%% file: B2G-21-003_temp.tex
\pdfoutput=1
\documentclass[11pt,twoside,a4paper,cmspaper,final,collab]{cms-tdr}
\def\svnVersion{0c3a6d3}\def\svnDate{2023/05/18}\def\cmsCernNoTag{CERN-EP-2022-034}\def\cmsCernDate{\today}\def\cmsMessage{Published in Physics Letters B as \href{http://dx.doi.org/10.1016/j.physletb.2022.137392}{\doi{10.1016/j.physletb.2022.137392}.}}
\begin{document}\cmsNoteHeader{B2G-21-003}

\newlength\cmsFigWidth
\ifthenelse{\boolean{cms@external}}{\setlength\cmsFigWidth{0.85\columnwidth}}{\setlength\cmsFigWidth{0.4\textwidth}}
\ifthenelse{\boolean{cms@external}}{\providecommand{\cmsLeft}{upper\xspace}}{\providecommand{\cmsLeft}{left\xspace}}
\ifthenelse{\boolean{cms@external}}{\providecommand{\cmsRight}{lower\xspace}}{\providecommand{\cmsRight}{right\xspace}}
\providecommand{\cmsTable}[1]{\resizebox{\textwidth}{!}{#1}}

\DeclareRobustCommand{\PX}{{\HepParticle{X}{}{}}\Xspace}
\newcommand{\PY}{{\HepParticle{Y}{}{}}\Xspace}
\newcommand{\PR}{{\HepParticle{R}{}{}}\Xspace}
\newcommand{\ttjets}{\ttbar{}+jets\xspace}
\newcommand{\XHbbYbb}{\ensuremath{\PX\to\PY\PH\to\bbbar\bbbar}\xspace}
\newcommand{\Hbb}{\ensuremath{\PH\to\bbbar}\xspace}
\newcommand{\gbb}{\ensuremath{\Pg\to\bbbar}\xspace}
\newcommand{\Ybb}{\ensuremath{\PY\to\bbbar}\xspace}
\newcommand{\HYbb}{\ensuremath{\PY/\PH\to\bbbar}\xspace}
\newcommand{\Rbb}{\ensuremath{\PR\to\bbbar}\xspace}
\newcommand{\XHY}{\ensuremath{\PX\to\PY\PH}\xspace}
\newcommand{\fb}{\ensuremath{\unit{fb}}\xspace}

\newcommand{\MX}{\ensuremath{M_{\PX}}\xspace}
\newcommand{\MY}{\ensuremath{M_{\PY}}\xspace}
\newcommand{\MJJ}{\ensuremath{M_{\mathrm{JJ}}}\xspace}
\newcommand{\MJY}{\ensuremath{M_{\mathrm{J}}^{\PY}}\xspace}
\newcommand{\MJH}{\ensuremath{M_{\mathrm{J}}^{\PH}}\xspace}
\newcommand{\MJt}{\ensuremath{M_{\mathrm{J}}^{\PQt}}\xspace}
\newcommand{\Rpf}{\ensuremath{R_{\mathrm{P/F}}}\xspace}
\newcommand{\Rpfinit}{\ensuremath{R_{\mathrm{P/F}}^{\text{init}}}\xspace}
\newcommand{\Rratio}{\ensuremath{R_{\text{ratio}}}\xspace}

\newcommand{\sigmaobs}{\ensuremath{\sigma_{\text{obs}}}\xspace}
\newcommand{\sigmaexp}{\ensuremath{\sigma_{\text{exp}}}\xspace}
\newcommand{\NNPDF}{\ensuremath{\textsc{NNPDF}\xspace}}

\newcommand{\intLumi}{\ensuremath{138\fbinv}\xspace}

\cmsNoteHeader{B2G-21-003}
\title{Search for a massive scalar resonance decaying to a light scalar and a Higgs boson in the four \texorpdfstring{\PQb}{b} quarks final state with boosted topology}

\date{\today}

\abstract{We search for new massive scalar particles \PX and \PY through the resonant process \XHbbYbb, where \PH is the standard model Higgs boson. Data from CERN LHC proton-proton collisions are used, collected at a centre-of-mass energy of 13\TeV in 2016--2018 and corresponding to an integrated luminosity of \intLumi. The search is performed in mass ranges of 0.9--4\TeV for \PX and 60--600\GeV for \PY, where both \PY and \PH are reconstructed as Lorentz-boosted single large-area jets. The results are interpreted in the context of the next-to-minimal supersymmetric standard model and also in an extension of the standard model with two additional singlet scalar fields. The 95\% confidence level upper limits for the production cross section vary between 0.1 and 150\fb depending on the X and Y masses, and represent a significant improvement over results from previous searches.}

\hypersetup{
pdfauthor={CMS Collaboration},%
pdftitle={Search for a massive scalar resonance decaying to a light scalar and a Higgs boson in the four b quarks final state with boosted topology},%
pdfsubject={CMS},%
pdfkeywords={LHC, CMS, NMSSM, Higgs, scalar, boosted jets}}

\maketitle 

\section{Introduction\label{sec:Introduction}}

The discovery of a Higgs boson (\PH) of mass 125\GeV~\cite{HiggsDiscoveryAtlas,HiggsDiscoveryCMS,CMSHiggsLongPaper} at the CERN LHC validated the Brout--Englert--Higgs mechanism~\cite{Englert:1964et,Higgs:1964ia,Higgs:1964pj,Guralnik:1964eu,Higgs:1966ev,Kibble:1967sv} of the standard model (SM), yet raised questions of its viability at higher energy scales~\cite{Buttazzo:2013uya,Hamada:2012bp,Jones:2013aua,Bednyakov:2015sca}. 
Besides, empirical observations such as the measurements of the neutrino masses and the baryon asymmetry in the universe are inconsistent with SM expectations. Beyond the standard model (BSM) theories, including those invoking supersymmetry~\cite{WZSUSY} or extra dimensions~\cite{RandallWED}, seek to address many of the shortcomings of the SM.
No BSM phenomena have been observed at the LHC. However, there are unexplored BSM parameter spaces, among which are areas of the scalar sector, the topic of this search.

The minimal supersymmetric extension of the SM (MSSM)~\cite{Fayet:1974pd,Fayet:1977yc} postulates two complex scalar field doublets with SU(2) gauge symmetry. The next-to-minimal model (NMSSM)~\cite{EllwangerNMSSM1,ManiatisNMSSM2} was proposed to solve the MSSM's ``unnaturalness problem''~\cite{KimMSSMmuProb}, where the Higgs boson mass parameter is many orders of magnitude smaller than the Planck scale. In the NMSSM, an extra complex scalar field gives a total of seven Higgs bosons: three neutral (one would be associated with \PH) and two charged scalar particles, as well as two neutral pseudoscalars. Searches for a heavier scalar \PX decaying to SM particles~\cite{ATLASHtotautau2020,CMSHtautau2018,Sirunyan:2684750} have set a lower limit on its mass at $\MX = 1.5\,(1.0)\TeV$ for $\tan\beta = 21\,(8)$~\cite{ATLASHtotautau2020}, where $\tan\beta$ is the ratio of the vacuum expectation values (VEVs) of the two Higgs doublets. The NMSSM favours low $\tan\beta$, where the current \MX bounds are the weakest.

In the NMSSM, the neutral scalar production cross sections may be suppressed because of their small couplings to SM fermions~\cite{EllwangerNMSSM1}. Enhanced ``Higgs-to-Higgs" decays are then possible, such as $\PX \to \PY\PH$~\cite{Djouadi:2008uw,Ellwanger:2017skc}, \PY being the lighter scalar. Within the NMSSM, the largest branching fractions for both \PH and \PY (for \PY mass \MY less than twice that of the top quark \PQt) are to a \PQb quark-antiquark pair, giving the final state $\PX \to \PY\PH \to \bbbar\bbbar$. For higher \MY values the $\PY \to \bbbar$ branching fraction is $\sim$10\%~\cite{Ellwanger:2017skc,Ellwanger2022}. The second dominant process is $\PX \to \PY\PH \to \tau\tau\bbbar$, which has been excluded~\cite{CMSXtoHYtotautaubb2021} for $0.4<\MX<0.6\TeV$ and $50<\MY<200\GeV$, for specific values of the parameters of the model. 

Another interesting model of new physics that motivates this search is the two-real-scalar-singlet extension of the SM (TRSM)~\cite{Robens:2019kga}, which introduces two additional scalar fields. 
This simplified model, onto which more complicated theories can be mapped, has nine degrees of freedom: the masses and VEVs of the three scalar fields, and three mixing angles.
In the scenario where all three VEVs are non-zero, the three fields give rise to three massive scalars, one of which can be associated with the \PH boson. 
Depending on their masses and mixing angles, the heaviest scalar can decay to the two lighter scalars. These in turn can decay to SM particles with mass-dependent branching fractions. 

This Letter describes the search for two new scalar particles, \PX and \PY, the former being more massive and decaying through \XHY. The search uses LHC proton-proton ($\Pp\Pp$) collision data collected by the CMS experiment in 2016--2018 corresponding to a total integrated luminosity of \intLumi~\cite{Sirunyan:2759951,CMS-PAS-LUM-17-004,CMS-PAS-LUM-18-002}. The masses of the scalar particles satisfy $\MX>\MY+\MH$; \PY may be lighter or heavier than \PH and both \PY and \PH decay to $\bbbar$. The search is generic, and \PX and \PY can be associated with the particles predicted in the NMSSM or the TRSM, which are both mentioned above.

This search focuses on the kinematic region where \MX is sufficiently larger than both \MY and $\MH$ such that \PY and \PH carry considerable momenta and therefore their decay products, \ie the $\bbbar$ pairs, are highly collimated. We explore the mass ranges $0.9<\MX<4\TeV$ and $60<\MY<600\GeV$, complementing the $\PX \to \PY\PH \to \tau\tau\bbbar$ search~\cite{CMSXtoHYtotautaubb2021}. In the high-momentum kinematic regime, special techniques are used to reconstruct the final states containing the collimated $\bbbar$ pairs, in order to increase the signal sensitivity well beyond that covered by previous searches~\cite{CMSXtoHYtotautaubb2021}. 

Tabulated results for this analysis are provided in HEPData~\cite{hepdata}.

\section{The CMS detector and event reconstruction}
The central feature of the CMS apparatus is a superconducting solenoid of 6\unit{m} internal diameter, providing a magnetic field of 3.8\unit{T}. Within the solenoid volume are a silicon pixel and strip tracker, a lead tungstate crystal electromagnetic calorimeter (ECAL), and a brass and scintillator hadron calorimeter (HCAL), each composed of a barrel and two endcap sections. Forward calorimeters extend the pseudorapidity ($\eta$) coverage provided by the barrel and endcap detectors. Muons are measured in gas-ionization detectors embedded in the steel flux-return yoke outside the solenoid. A more detailed description of the CMS detector, together with a definition of the coordinate system used and the relevant kinematic variables, can be found in Ref.~\cite{CMSJINST}. Events of interest are selected using a two-tiered trigger system. The first level, composed of custom hardware processors, uses information from the calorimeters and muon detectors to select events at a rate of around 100\unit{kHz} within a fixed latency of about 4\mus~\cite{CMSL1}. The second level, known as the high-level trigger (HLT), consists of a farm of processors running a version of the full event reconstruction software optimized for fast processing, and reduces the event rate to around 1\unit{kHz} before data storage~\cite{CMSHLT}.

The primary vertex is taken to be the vertex corresponding to the hardest scattering in the event, evaluated using tracking information alone, as described in Section 9.4.1 of Ref.~\cite{CMS-TDR-15-02}.

A particle-flow algorithm (PF)~\cite{CMSPF} aims to reconstruct and identify each individual particle in an event, with an optimized combination of information from the various elements of the CMS detector. The photon energy is obtained from the ECAL measurements. The energy of electrons is determined from a combination of the electron momentum at the primary interaction vertex as determined by the tracker, the energy of the corresponding ECAL cluster, and the energy sum of all bremsstrahlung photons spatially compatible with originating from the electron track. The energy of muons is obtained from the curvature of the corresponding track. The energy of charged hadrons is determined from a combination of their momentum measured in the tracker and the matching ECAL and HCAL energy deposits, corrected for the response function of the calorimeters to hadronic showers. Finally, the energy of neutral hadrons is obtained from the corresponding corrected ECAL and HCAL energies.

Jets are clustered from PF candidates using the anti-\kt algorithm~\cite{Cacciari:2008gp,Cacciari:2011ma} with a distance parameter of either 0.4 (AK4 jets) or 0.8 (AK8 jets). 
The jet momentum is defined as the vectorial sum of all particle momenta in a jet, and is found from simulation to be, on average, within 5--10\% of the true momentum over the whole transverse momentum (\pt) spectrum and detector acceptance~\cite{CMS:2016lmd}. 
Additional $\Pp\Pp$ interactions within the same or nearby bunch crossings (pileup), averaging 23--32 in 2016--2018, can contribute additional tracks and calorimetric energy depositions to the jet momentum. 
The effect of pileup is mitigated using the charged-hadron subtraction (CHS) algorithm~\cite{Sirunyan:2020foa}, whereby charged particles identified to be originating from pileup vertices are discarded and an offset correction is applied to correct for remaining contributions. 
Jet energy corrections are derived from simulation and data to bring the measured response of jets to that of particle level jets on average~\cite{CMS:2016lmd}.

For AK8 jets, masses are computed after applying grooming~\cite{SalamJetography} techniques, which remove wide-angled soft and collinear radiation from the jets, in order to mitigate the effects of contamination from initial state radiation, the underlying event, and multiple hadron scattering. The trimming algorithm~\cite{Krohn:2009th} uses a subjet size parameter of 0.3 and a radiation fraction parameter $z=0.1$, which determines the minimum \pt fraction that the reclustered jet constituents need to have in order not to be removed. The mass of the resultant jet is referred to as its ``trimmed mass''. The ``soft-drop mass'' of the jet is obtained by applying the soft-drop algorithm ~\cite{Dasgupta:2013ihk,Larkoski:2014wba}. Here it is obtained using a value $z=0.1$ for the radiation fraction parameter. The angular exponent parameter is set as $\beta=0$, so there is no dependence of the \pt fraction threshold on the distance between jet constituents.

In case of the soft-drop algorithm, the pileup per particle identification (PUPPI)~\cite{Sirunyan:2020foa,Bertolini:2014bba} algorithm is used to mitigate the effect of pileup on AK8 jets. In PUPPI, the treatment of charged particles is similar to that in CHS. A weight between zero and one is assigned to neutral particles, larger values indicating higher probability of originating from the primary interaction vertex. The jet mass is computed from the weighted sum of the constituent four-momenta.

The missing transverse momentum vector (\ptvecmiss) is computed as the negative vector sum of the transverse momenta of all the PF candidates in an event, and its magnitude is denoted as \ptmiss~\cite{CMSMET}. The \ptvecmiss is modified to account for corrections to the energy scale of the reconstructed jets in the event. 

\section{Signal and background processes\label{sec:EvtSim}}

Monte Carlo simulations of the signal process \XHbbYbb, with a width of 1\MeV for all the three scalars, are generated at leading order (LO) using the {\MGvATNLO{}2.6.5}~\cite{Madgraph} event generator. The NMSSM model~\cite{NMSSMufo,UFO} is used to produce the simulated samples. However, the kinematic parameters are model-independent, enabling the results to be interpreted using other BSM scenarios.

The two main backgrounds are \ttjets events, where the top quarks decay hadronically, and events with jets arising purely from SM quantum chromodynamics (QCD) interactions (multijet events). 
Other sources of background like single top quark production, and Higgs boson production in association with a top quark pair or a \PW or \PZ boson are found to have negligible contributions. 

The \ttjets events with hadronic top quark decays are modelled using {\POWHEG{}2.0}~\cite{POWHEG,POWHEG_Frixione,POWHEGBOX,POWHEG_Nason}, at next-to-leading order (NLO). A sample of semileptonic \ttbar decays, with one of the top quarks decaying via $\PQt \to \PW\PQb \to \ell\nu\PQb$, $\ell$ being a lepton (electron or muon), is also simulated using the same configuration. These events are used in dedicated \ttbar enriched control regions to derive additional data-to-simulation correction factors. The simulated \ttjets event yields are scaled using a cross section of $832^{+46}_{-52}\unit{pb}$, calculated at next-to-next-to-leading order (NNLO) in QCD with soft gluon resummation at next-to-next-to-leading logarithmic precision~\cite{Czakon:2011xx}. The QCD multijet event samples, containing two to four jets, are simulated at LO using the {\MGvATNLO{}} event generator and are used to develop the tools for the analysis. However, this background is estimated using data-driven techniques.

The signal and semileptonic \ttjets samples are generated using the {\NNPDF{}3.1}~\cite{NNPDF31} NNLO parton distribution functions (PDFs) from the {\textsc{lhapdf}{}6} PDF library~\cite{Buckley:2014ana}. The hadronic \ttjets samples are generated using  {\NNPDF{}3.0}~\cite{NNPDF30} NLO for 2016 and {\NNPDF{}3.1} NNLO for 2017 and 2018 simulation. The multijet background samples are generated using {\NNPDF{}3.0} LO for the 2016 and {\NNPDF{}3.1} NNLO for the 2017 and 2018 simulation. 

The showering and hadronization of partons are simulated with {\PYTHIA{}8.226}~\cite{Pythia8p2} for 2016 and {\PYTHIA{}8.240} for 2017 and 2018 samples. The jet-to-parton matching for all LO samples, \ie the signal and the multijets background, use the MLM~\cite{MLM} scheme. The \textsc{CP5} tune~\cite{CP5} is used for all samples, except for the 2016 \ttbar and multijet samples, which use \textsc{CUETP8M2T4}~\cite{CUETP8M2T4} and \textsc{CUETP8M1}~\cite{CUETP8M1} tunes, respectively. 

All generated events are processed through a simulation of the CMS detector based on \GEANTfour~\cite{GEANT}. The effects of pileup are modelled assuming a total inelastic $\Pp\Pp$ cross section of {69.2\unit{mb}}~\cite{CMSLHCppXsec}. All simulated event samples are weighted to match the distribution of the expected pileup profile of the data.

\section{Event selection\label{sec:EvtSel}}

The events are selected in two mutually-exclusive categories: an ``all-jets'' event sample containing only jets, and a ``jets+lepton'' sample, containing a lepton (electron or muon). The latter serves to derive corrections to the simulated \ttjets background, in order to match the expectations in the data.

\subsection{All-jets event selection\label{ss:all-jets}}
A set of triggers based on requirements on jet properties are used for online event selection in the all-jets category.

One trigger criterion required a single AK8 jet with $\pt>450$ or 500\GeV in 2016 and in 2017--2018, respectively. A second trigger required that the scalar sum (\HT) of the \pt of all AK4 jets with $\pt>30\GeV$ and $\abs{\eta}<2.5$ should be greater than 800 or 900\GeV in 2016, depending on the LHC beam instantaneous luminosity. In 2017--2018, $\HT>1050\GeV$ was required. 

The third trigger algorithm used required an AK8 jet with a trimmed mass $>$30\GeV along with $\pt>360\GeV$ (in 2016). In 2017--2018, the AK8 jet \pt threshold in this trigger was raised to 400 or 420\GeV, depending on the LHC beam instantaneous luminosity, keeping the same trimmed mass criterion. The fourth trigger required $\HT > 650$ or 700\GeV (in 2016) and $\HT > 800\GeV$ (in 2017--2018), together with an AK8 jet having a trimmed mass $>$50\GeV. 

In addition to the above, three trigger algorithms were used in 2016 only, with the following criteria: (1) two AK8 jets with $\pt > 280$ and $>$200\GeV with one of them having a trimmed mass $>$30\GeV; (2) having the same requirements as (1) and with one of the AK8 jets passing a loose \PQb tagging criterion using the ``combined secondary vertex'' algorithm~\cite{CMSBTagging2018} (with efficiency $\approx$ 81\%); (3) $\HT\ge650\GeV$ with a pair of AK4 jets having an invariant mass $>$900\GeV with their pseudorapidity separation $\abs{\Delta\eta} < 1.5$.

The combined logical OR of all the triggers improves the overall trigger efficiency, particularly for signals with low values of \MX. 

Events in the offline all-jets selection are required to have at least two AK8 jets with $\pt>350\,(450)\GeV$ and $\abs{\eta} < 2.4\,(2.5)$ for 2016 (2017--2018). The higher \pt requirement in 2017--2018 reflects the higher trigger thresholds and ensures a trigger efficiency close to 100\%. 
The AK8 jet pairs in multijet backgrounds tend to have a larger separation in pseudorapidity than the signal, for a given \MJJ range, and therefore a selection $\abs{\Delta\eta}<1.3$ is used to reduce such backgrounds.

The two leading-\pt jets are considered for \Hbb and \Ybb candidates. An \Hbb candidate or an ``\PH jet'' is a jet whose soft-drop mass is $110<\MJH<140\GeV$. The second jet is designated as the \Ybb candidate or the ``\PY jet'' if its soft-drop mass satisfies $\MJY > 60\GeV$. When both AK8 jets satisfy the first mass requirement, the \PY jet is chosen at random. Events without either an \PH or a \PY jet are rejected. The mass of the \PY jet and the invariant mass of the \PH and \PY jets are used to isolate the signal with approximately 15\% and 9\% resolution in \MJY and \MJJ, respectively.

The all-jets event category trigger efficiency is measured in the data requiring a single AK4 jet with $\pt > 260\GeV$ by applying the above offline selection, and counting the number of events passing the trigger selection. It is found to be between 94 and 100\%. Simulated events are weighted by this efficiency as a function of the invariant mass of the two leading-\pt AK8 jets in the event, \MJJ.

A graph convolutional neural network algorithm, ParticleNet~\cite{PNet_HQu_LGouskos}, is employed to discriminate the decays of a boosted massive particle \PR, which could be an \PH boson or a \PY resonance, to a pair of \PQb quarks against a background of other jets, using the properties of the jet PF constituents as features. As with all heavy-flavour jet classifiers, displaced tracks and vertices are the most important features.
The multiclassifier ParticleNet algorithm outputs several variables, each in the range 0--1, and each of which can be interpreted as the probability of a jet having originated from a certain decay, such as from a massive resonance \Rbb ($P(\Rbb)$) or from a light-flavoured quark or a gluon ($P(\text{QCD})$). In this analysis, the ParticleNet score is defined as $P(\Rbb)/(P(\Rbb) + P(\text{QCD}))$, where $P(\Rbb)$ is a unified score for jets originating from \PH or \PY decays.

The ParticleNet algorithm is trained~\cite{CMS-DP-2020-002} on AK8 jets using as the signal simulated Lorentz-boosted spin-0 particles decaying to a pair of \PQb quarks, with a wide range of masses. The QCD multijet samples are used for the background. The wide signal mass range in the training sample ensures that the background rejection rate is decorrelated from the mass of the jet~\cite{CMS-DP-2020-002}. As a consequence, background enriched regions can be defined using low ParticleNet scores on jets that have the same mass spectra as that of the background in the signal region. An accurate background model can therefore be developed. 

\begin{figure}[ht]
  \centering
    \includegraphics[width=0.49\textwidth]{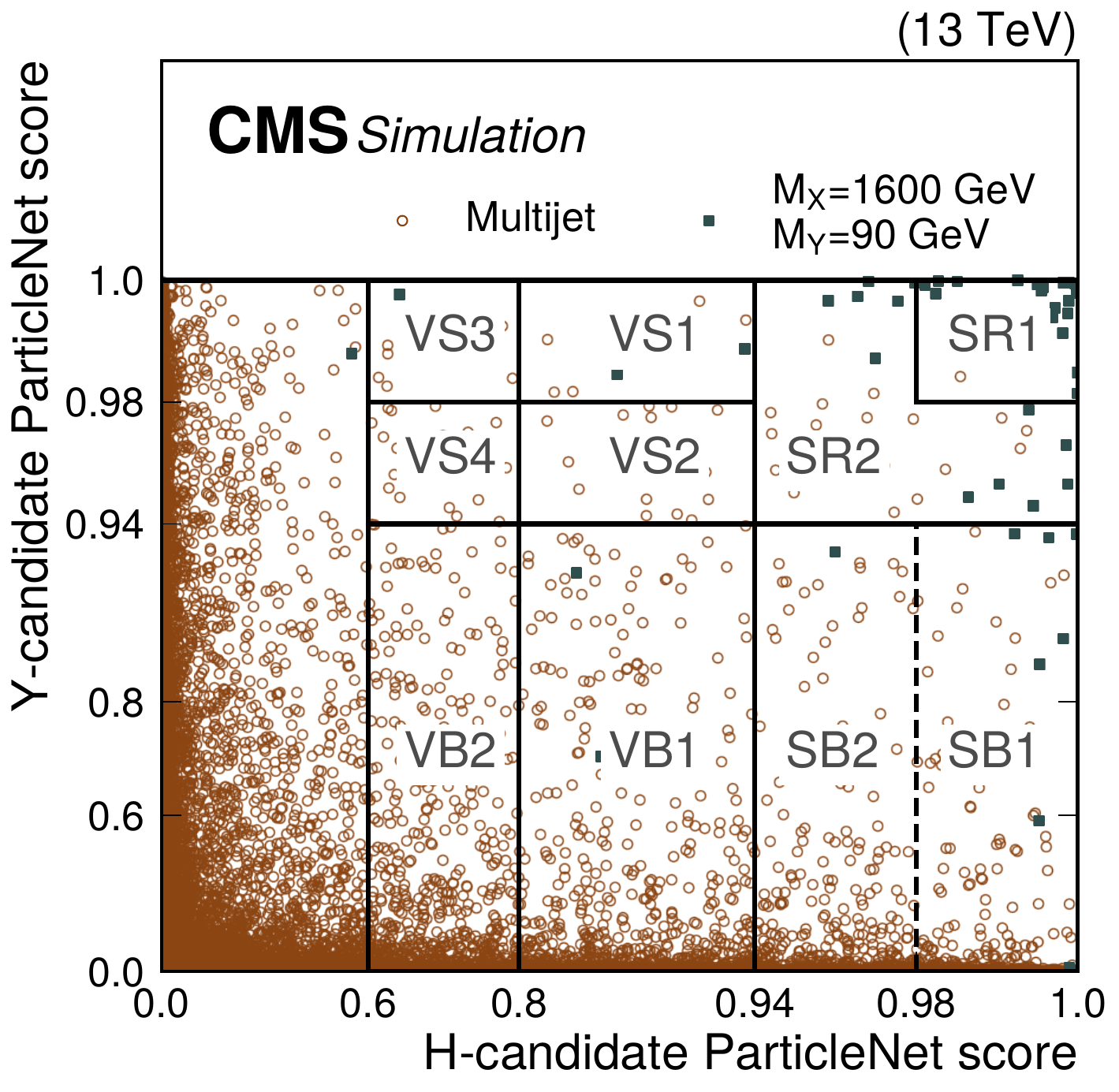} 
  \caption{Simulated ParticleNet score distributions of the \PH and the \PY candidate jets for a signal with $\MX=1600\GeV$ and $\MY=90\GeV$ (filled squares) and the multijets background (open circles). The grid lines show the different event categories defined using the ParticleNet scores of the two jets. A description of the regions is given in the text.}
  \label{fig:ParticleNet_regions}
\end{figure}

The ParticleNet scores used for selecting the \Hbb and the \Ybb candidates (``signal jets'') are either $>$0.98 (tight requirement) or $>$0.94 (loose requirement). Depending on the jet \pt, the former has an efficiency of 62--72\% and a misidentification rate of 0.45\%, while the latter has an efficiency of 80--85\% and a misidentification rate of 1\%. 

The efficiency of the ParticleNet classifier is calibrated in data using a sample of jets originating from fragmentation of a gluon to $\bbbar$ (\gbb), which are similar to \Hbb and \Ybb jets. Such jets are selected from the data using a boosted decision tree (BDT) classifier, such that their ensemble ParticleNet score resembles that of \PY and \PH jets. Using simulated multijet events, the BDT is trained to separate \gbb jets from jets of other flavours. A systematic uncertainty is assigned to account for different possible choices of such jets. The measurements give a data-to-simulation correction factor of 0.9--1.4 for the ParticleNet selection efficiencies, depending on the jet \pt and data-taking year.

The ParticleNet scores of the \PH and \PY jets are used to classify events into either signal, sideband, or validation categories. A layout of the different regions is shown in Fig.~\ref{fig:ParticleNet_regions}.

Two signal regions are defined using the tight and the loose ParticleNet scores (Fig.~\ref{fig:ParticleNet_regions}). The ``signal region 1" (SR1) and the ``signal region 2" (SR2) are statistically exclusive. SR1 has a higher signal-to-background ratio and is thus more sensitive to the presence of signal. However, the SR2 improves the sensitivity for signal mass points with low background by increasing the signal efficiency. 

Corresponding to the two signal regions, two ``sideband regions'' are defined for estimating the multijet background from data. They are labelled as ``Sideband 1" (SB1) and ``Sideband 2" (SB2) in Fig.~\ref{fig:ParticleNet_regions}. The SB2 region includes SB1 in order to provide better sideband region characteristics for estimating the multijets background in their respective signal regions.

In addition, six ``validation regions'' are used to validate the background estimation method. They are grouped into two sets of three regions: VS1, VS2, VB1, and VS3, VS4, VB2 as shown in Fig.~\ref{fig:ParticleNet_regions}. The labels VS and VB stand for ``signal-like'' and ``background-like'' validation regions, respectively. All these regions are enriched in QCD multijet events, and have much smaller signal-to-background ratios than the signal regions. 

The signal selection efficiencies range from 1.7\% to 12.6\% in SR1 and 1.3\% to 5.6\% in SR2. Based on simulation, the background composition is about an equal proportion of \ttjets and QCD multijets in the signal regions and in the corresponding validation regions, VS1--VS4. However, the sideband and validation sideband regions are composed $\approx$ 90\% of multijet events.

\subsection{Jets+lepton event selection\label{ss:jetslepton}}

The triggers in the jets+lepton category required events to have either an isolated muon of $\pt > 24$ or 27\GeV; an isolated electron having $\pt > 27$, 32, or 35\GeV, or a photon with $\pt>175$ or 200\GeV. The thresholds changed between data-taking years. The jets+lepton event trigger efficiencies are measured in a sample of $\PZ\to\ell\ell$ events and are found to be close to 100\%. 

Offline selection requires the events to have a lepton with $\pt>40\GeV$ and $\abs{\eta}<2.4$. Tight identification and isolation criteria are used for electrons~\cite{CMS:2015xaf} and muons~\cite{CMS:2018rym}. An AK4 jet, corrected for pileup using charged hadron subtraction~\cite{Sirunyan:2020foa} and tagged as originating from a bottom quark (\PQb-tagged) using the DeepJet algorithm~\cite{DeepJet}, is required to be close to the lepton. The criterion is $\Delta R(\text{lepton, jet})<1.5$, where $\Delta R(1,2) \equiv \sqrt[]{\smash[b]{(\eta_{1}-\eta_{2})^{2}+(\varphi_{1}-\varphi_{2})^{2}}}$ is the distance between two objects in the pseudorapidity--azimuthal angle plane. 

The loose DeepJet working point, with a mistag rate of 10\% and approximately 90\% efficiency, is used. The DeepJet score distributions of the AK4 jets are corrected using a weight extracted from measurements in the data~\cite{CMSBTagging2018}. Requirements of $\ptmiss>60\GeV$ and $\HT>500\GeV$ are imposed. The lepton, \ptmiss, and the \PQb-tagged jet provide the signature of the leptonic decay of a top quark. A hadronically decaying top quark candidate is reconstructed from an AK8 jet with $\pt>350\,(450)\GeV$ and $\abs{\eta}<2.4\,(2.5)$ for 2016 (2017--2018), a soft-drop mass $>$60\GeV, and satisfying $\Delta R (\text{lepton, AK8 jet})>2$. Events in the jets+lepton category, which has a purity of $>$90\%, are split into two regions based on whether the AK8 jet passes the tight or loose ParticleNet scores. Two separate correction factors are derived, one each for SR1 and SR2.

\section{Background estimation}\label{sec:bkgEst}
The analysis searches for a narrow signal in the 2-dimensional plane spanned by \MJJ and \MJY. The two-dimensional (\MJJ, \MJY) distributions of the multijet events are estimated using a pass-to-fail ratio method, described in the following paragraphs. The simulated \ttjets event distributions are corrected by fitting the top quark jet mass \MJt distributions to the data in the jets+lepton regions.

The multijet background is estimated for the three data-taking years combined. First, transfer functions, \Rpf, are defined as the ratio of event distributions in the (\MJJ, \MJY) plane in the signal regions to those in the sidebands, SR1-to-SB1 and SR2-to-SB2. These are a priori unknown, and are determined from the fit of signal and background distributions to the data.

An initial estimate \Rpfinit is made using the first set of validation regions, using the data and correcting for the simulated \ttjets component: VS1-to-VB1 and VS2-to-VB1. With the definition $\Rpf \equiv \Rpfinit\Rratio$, only the correction function \Rratio needs to be determined directly from the fit to the data. The validation regions provide a good estimate of \Rpf, because the pass-to-fail event ratios SR1-to-SB1 and SR2-to-SB2 are close to VS1-to-VB1 and VS2-to-VB1, as borne out in simulations. The values of \Rratio are therefore of order unity and lead to stability of the fit of signal and background models to the data.

The values of \Rpfinit, closely related to the loose and tight misidentification rates of the ParticleNet tagger, are determined as functions of \MJY only. The 1-dimensional modelling reduces the statistical uncertainties in the \Rpfinit.  A quadratic function is found to be the best model. Furthermore, the \Rpf dependence on \MJJ is weaker and is modelled through the \Rratio, determined directly from the fit to the data in the signal regions.

The form of the \Rratio is chosen to be a product of two polynomials in \MJY and \MJJ, whose parameters are determined from a simultaneous fit of the binned signal and background distributions to the data in SR1, SR2, SB1, and SB2. A variable bin width over the (\MJJ, \MJY) plane  was chosen to correspond to the signal resolution while ensuring that there were no zero-event bins in the sideband regions.

A Fisher's F-test~\cite{Fisher} is used to determine the minimum polynomial order necessary and sufficient for the model. Starting from polynomials of order one in both \MJJ and \MJY, terms are added until no significant improvement is observed. The F-test shows that linear functions in both \MJJ and \MJY are favoured at 95\% confidence level (\CL). The two \Rratio values range from 0.4 to 2.9 over the whole (\MJJ, \MJY) plane.

The simulated \ttjets event distributions in (\MJJ, \MJY) for the signal regions are corrected for their shape and yield using the jets+lepton event category, which is highly enriched in this background. The AK8 jets from top quark decays fall into three categories, depending on the top quark boost. A high enough boost may result in a fully merged $\PQt \to \PW\PQb \to \PQq\PAQq'\PQb$ decay, labelled as a $\PQb\PQq\PQq$ jet. At moderate boosts, the $\PW \to \PQq\PAQq'$ may be merged to form a \PW jet with the \PQb quark forming its own jet. However, such events are nearly all eliminated in the event selection. Finally, one of the quarks from the \PW boson decay can merge with the \PQb quark to form a $\PQb\PQq$ jet. Unmerged jets and other combinatorial backgrounds constitute a small fraction outside these three categories.

The masses of the $\PQb\PQq\PQq$ and $\PQb\PQq$ jet components in the jets+lepton event category are fit to the data simultaneously with the all-jets event categories. These two mass distributions are scaled independently, with each being tied to the corresponding jet component from the \ttjets in SR1 and SR2. 
They are independent for the three years, giving six scales in total ranging from 0.79 to 1.35.

Two sets of cross-checks are performed for the background estimation method. The first is to predict the background in the validation regions VS1 and VS2 using the validation region VB1 as a sideband. The \Rpfinit are estimated from the ratios of events in the regions VS3 to VB2 and VS4 to VB2. The jets+lepton regions are treated as they would be for the true background estimation in the signal regions. Similar to the actual background estimation process, a Fisher's test is used to decide the polynomial order of the \Rratio function. Again, the most favoured form for \Rratio is found to be the product of linear functions along both \MJJ and in \MJY. A goodness-of-fit test confirmed the agreement between the data and the estimated background, with the p-value greater than 0.05. 

The second check uses generated toy data sets for SR1 and SR2. A toy QCD multijets background is first obtained for these regions by applying the \Rpf of VS1 and VS2, obtained in the first validation exercise, to SB1 and SB2, respectively. The toy multijets background is then combined with the \ttjets sample and different signal strengths to get the toy data sets. The test consists of comparing the estimated and true signal strengths after the full background estimation and signal extraction procedure. The test shows no bias in the estimated signal yields for a wide range of \MX and \MY. 

\section{Systematic uncertainties}\label{sec:sysUnc}

Several sources of systematic uncertainty affect the (\MJJ, \MJY) shapes and the yields of the signals and backgrounds. The impact of the systematic uncertainties is reported for a signal with $\MX=1.6\TeV$ and $\MY=150\GeV$.
\begin{itemize}
\item \textit{ParticleNet scale factor}: the uncertainty is 7--37\%, depending on the AK8 jet \pt, and affects the signal by 15\%.
\item \textit{Jet energy scale and resolution}: the uncertainties are applied to both AK4 and AK8 jets, and are fully correlated between the two sets of jets. The signal is affected by 5\%.
\item \textit{Jet mass scale}: this is modelled as a $\pm 5\%$ shift in the AK8 jet soft-drop mass. It is uncorrelated between the $\PQb\PQq\PQq$, the $\PQb\PQq$, and the signal jets. It affects the signal by 13\%. The jet mass scale uncertainty in the \ttjets background is reduced by including the jets+lepton control region.
\item \textit{Jet mass resolution}: simulated AK8 jet masses are smeared to match their distributions in the data, based on studies using Lorentz-boosted $\PW \to \PQq\PAQq'$ (\PW boson jets). The nominal simulated unsmeared jet mass resolution is taken as the downward uncertainty while applying a 20\% larger smear~\cite{Sirunyan:2020foa} is used to estimate the upward uncertainty in the AK8 jet mass resolution. The resultant impact on the signal yield is an uncertainty of 4\%. 
\end{itemize}

The following uncertainties affect only the backgrounds.
\begin{itemize}
\item \textit{\ttbar normalization}: the uncertainties in the $\PQb\PQq\PQq$ and $\PQb\PQq$ jet scale factors range from 6\% to 16\%.
\item \textit{Top quark \pt modelling in Monte Carlo simulations}: an uncertainty is assigned to the \ttjets simulation process~\cite{CMS-PAS-TOP-17-007}, resulting in a 2\% uncertainty in this background.
\item \textit{Multijets background uncertainty}: the uncertainty derives mainly from the uncertainty in the $\Rpfinit (\MJY)$, which is driven by the sample sizes in the sideband regions VS1, VS2, and VB1. It corresponds to a 7--11\% change in the multijet background yields.
\end{itemize}
Other systematic uncertainties with minor impact are the following.
\begin{itemize}
\item \textit{Trigger efficiency}: the difference between the jet energy scale at the HLT and in the offline reconstruction~\cite{HHbbbb2016Paper} is appreciable for $\MJJ<1100\GeV$, resulting in the trigger efficiency dropping below 100\%. An uncertainty equal to half of the difference between unity and the measured trigger efficiency is assigned. It is larger than the statistical uncertainty and is expected to cover jet energy scale uncertainties in the trigger efficiencies. Its maximum value is 3\%.
\item \textit{Trigger timing correction}: during the 2016--2017 data taking, a gradual shift in the timing of the inputs of the ECAL hardware level trigger in the region of $\abs{\eta} > 2$ caused a specific trigger inefficiency. To take this effect into account, a 2\% normalization uncertainty is applied to \ttbar events and signal for these years.
\item \textit{Integrated luminosity}: the uncertainty in the total Run 2 (2016--2018) integrated luminosity~\cite{CMS-PAS-LUM-17-004,CMS-PAS-LUM-18-002} is 1.6\%.
\item \textit{Pileup}: the value of the $\Pp\Pp$ total inelastic cross section that is used in the simulation of pileup events is varied upwards and downwards from its assumed value of {69.2\unit{mb}} by its uncertainty of 4.6\%~\cite{CMSLHCppXsec}.
\item \textit{PDF and scale uncertainties}: the impact of the PDF and the QCD factorization $\mu_{\mathrm{F}}$ and renormalization $\mu_{\mathrm{R}}$ scale uncertainties in the signal acceptance and selection is estimated to be 1\%. The former is derived using the {PDF4LHC} procedure~\cite{Butterworth:2015oua} and the {NNPDF3.1} PDF sets. The latter is evaluated by separately changing $\mu_{\mathrm{R}}$ and $\mu_{\mathrm{F}}$ in simulation by factors of 0.5 and 2.
\item \textit{Sample size of sideband regions}: the effects of the limited sizes of the SB1 and SB2 samples are included as statistical uncertainties in the multijets background predicted in SR1 and SR2, using the Barlow--Beeston Lite prescription~\cite{BarlowBeeston,BBLite}. These uncertainties are small compared to the uncertainties in \Rpfinit. 
\item \textit{Lepton ID and isolation efficiencies}: the data-to-simulation correction factors for the efficiencies have uncertainties that affect the event yields by 1--2\% in the jets+lepton selection.
\item \textit{AK4 jet \PQb-tagging data-to-simulation scale factor uncertainty}: this uncertainty amounts to about 2\% and affects the semileptonic \ttjets event yields.
\end{itemize}
All uncertainties affecting the signal and the \ttjets samples are uncorrelated among years, except those associated with the PDF choice, the renormalization and factorization scales, the pileup correction, the integrated luminosity, and the top quark \pt modelling.

\section{Results}\label{sec:results}

\begin{figure}[!htbp]
  \centering 
    \includegraphics[width=0.49\textwidth]{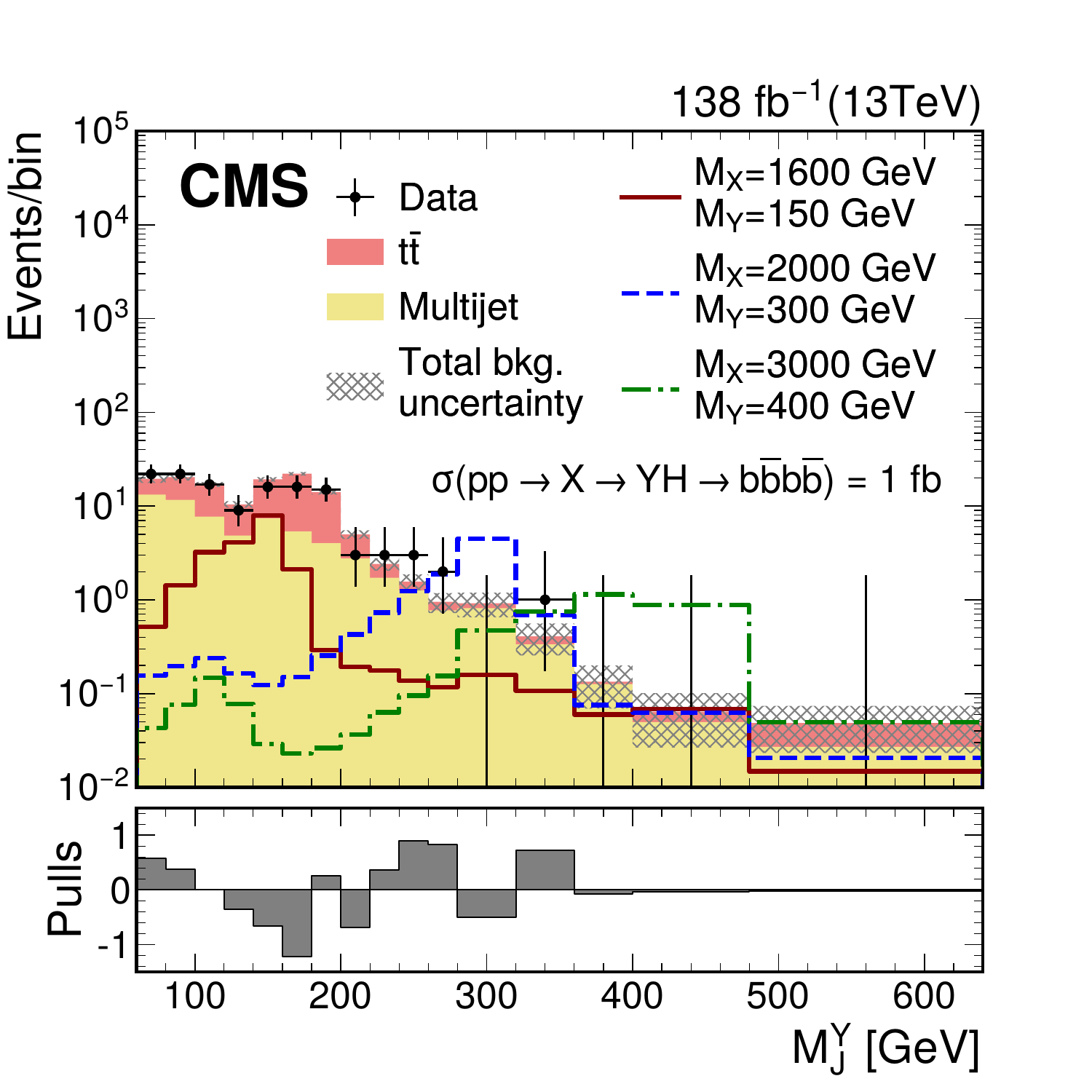} 
    \includegraphics[width=0.49\textwidth]{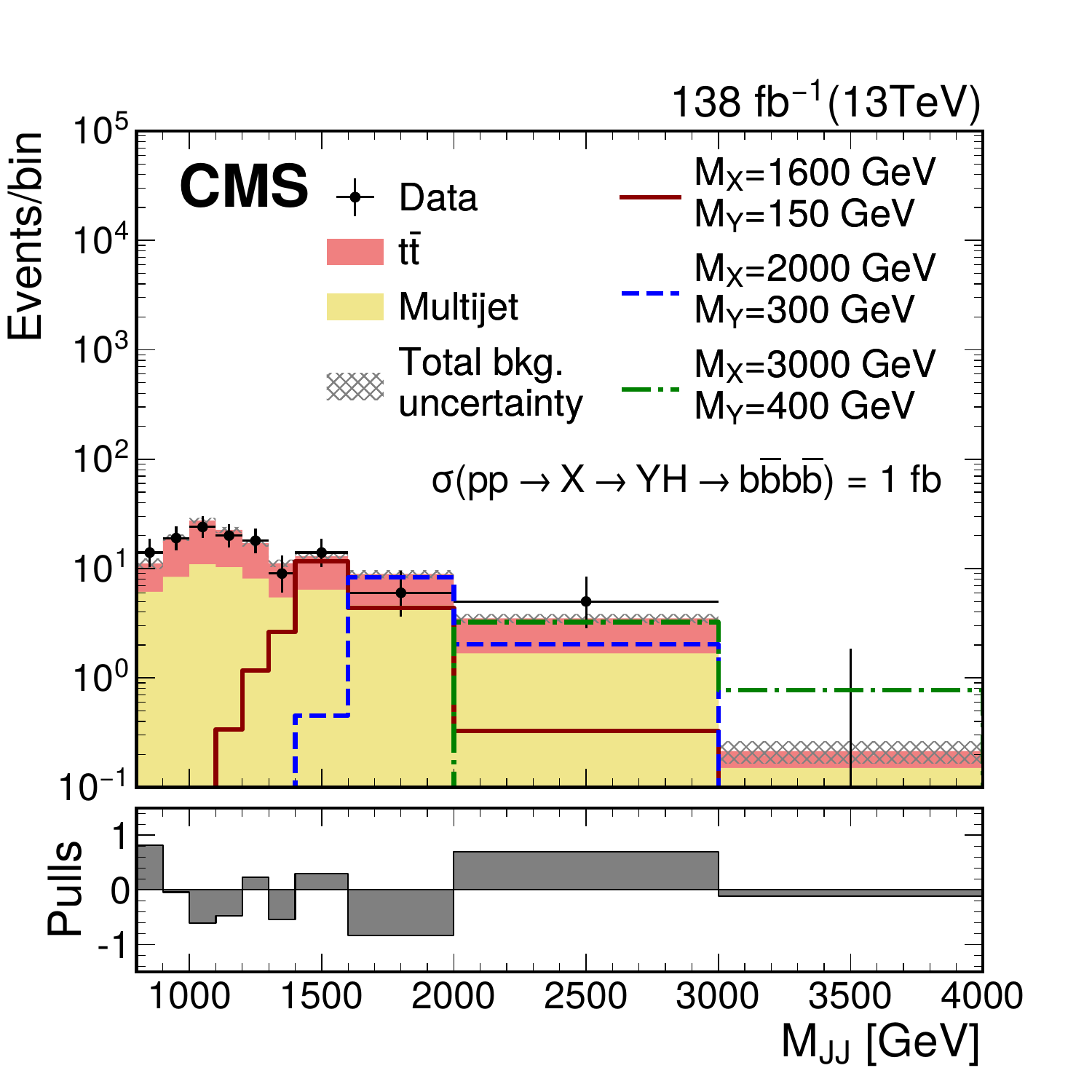} 
  \caption{The \MJY (\cmsLeft) and \MJJ (\cmsRight) distributions for the number of observed events (black markers) compared with the estimated backgrounds (filled histograms) and their uncertainties (hatched areas) in SR1. The distributions expected from the signal under three \MX and \MY hypotheses and assuming a cross section of 1\fb are also shown. The lower panels show the ``Pulls'' defined as $(\text{observed events}{-}\text{expected events})/\sqrt{\smash[b]{\sigmaobs^{2} - \sigmaexp^{2}}}$, where \sigmaobs and \sigmaexp are the statistical and total uncertainties in the observation and the background estimation, respectively. The minus sign accounts for the correlation between data and the data-driven estimation.}
  \label{fig:TT}
\end{figure}

\begin{figure}[!htbp]
  \centering 
    \includegraphics[width=0.49\textwidth]{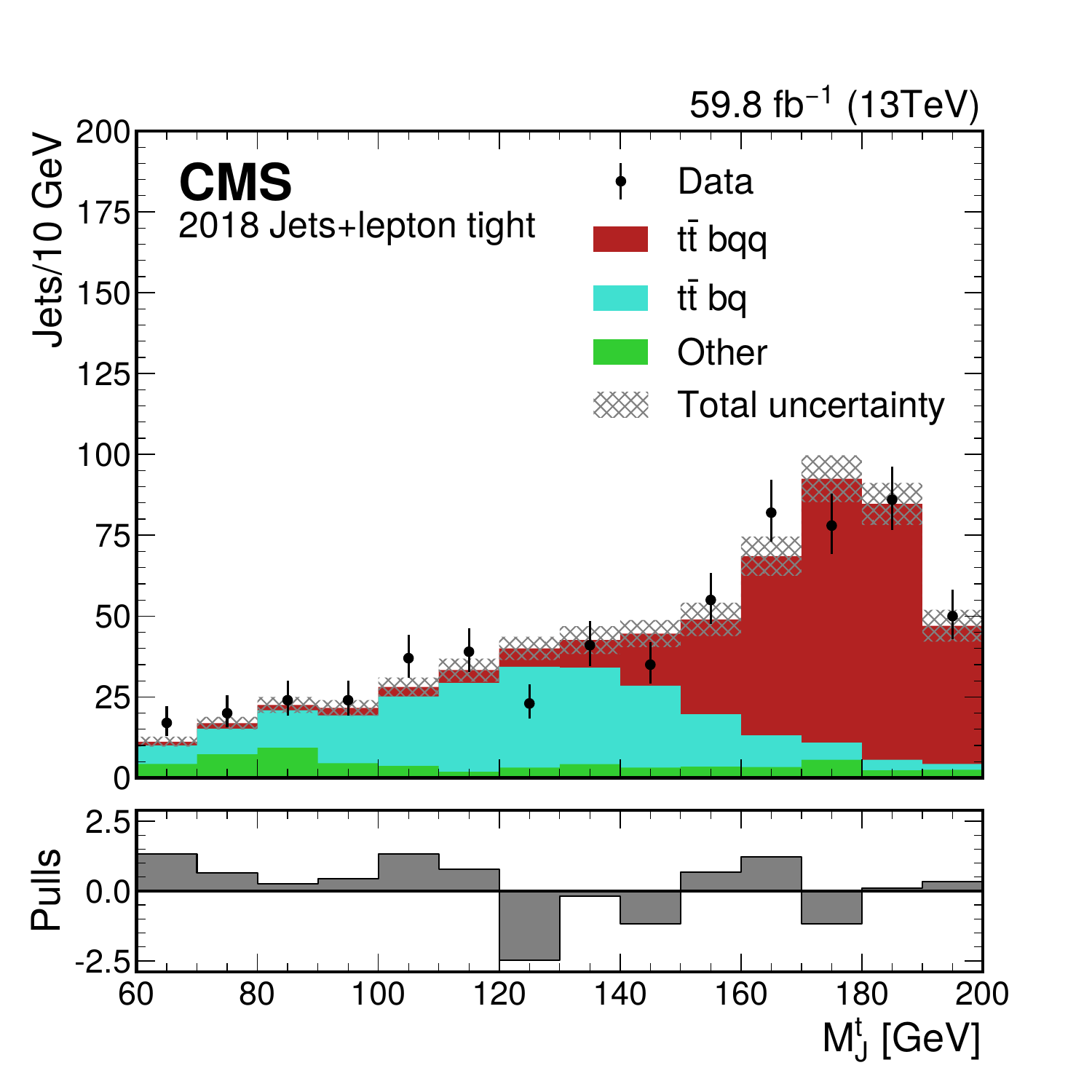} 
  \caption{The soft-drop mass distributions of the top quark candidate jets in the 2018 jets+lepton category, in the tight ParticleNet region, after the joint fit in the all-jets and jets+lepton categories. The observed data (black markers) and the post-fit estimate (filled histograms) are shown for the three jet categories. The lower panels show the ``Pulls'' defined as $(\text{observed events}{-}\text{expected events})/\sqrt{\smash[b]{\sigmaobs^{2} + \sigmaexp^{2}}}$, where \sigmaobs and \sigmaexp are the total uncertainties in the observation and the background estimation, respectively.}
  \label{fig:CR}
\end{figure}

The joint likelihood of the signal$+$background (\MJJ, \MJY) distributions in the all-jets regions (SR1 and SR2; SB1 and SB2), along with the \MJt distributions in the jets+lepton tight and loose regions is constructed. The binned signal (\MJJ, \MJY) distributions are extracted from  260 signal hypothesis simulations. A combined three-year multijets background component is used in the likelihood distribution to reduce the statistical uncertainty. However, the likelihood has three separate \ttjets and signal components, one for each data-taking year. For the three years, 2016--2018, the data distributions in the all-jets regions are added, while those for the jets+lepton regions are kept separate. The uncertainties are treated as nuisance parameters while fitting the signal+background models to the data. The distributions of the data, the post-fit background, and three representative signals in SR1 are shown in Fig.~\ref{fig:TT}. The fitted distributions in the jets+lepton tight region, used for determining the corrections to the simulated \ttjets events, are shown in Fig.~\ref{fig:CR} (for 2018).

The signal hypothesis with $\MX=1.6\TeV$ and $\MY=90\GeV$ gives the highest observed local significance of $3.1\sigma$, which becomes $0.7\sigma$ after accounting for the look-elsewhere effect~\cite{Vitells2011}. 
However, the excess is not apparent in Fig.~\ref{fig:TT}, which shows the separate 1-dimensional distributions of \MJJ and \MJY, integrated over the other variable. 
The estimated background is otherwise in agreement with the observed data. Upper limits on the signal cross section are calculated for various hypothesized values of \MX and \MY.

\begin{figure}[!htb]
  \centering 
  \includegraphics[width=0.49\textwidth]{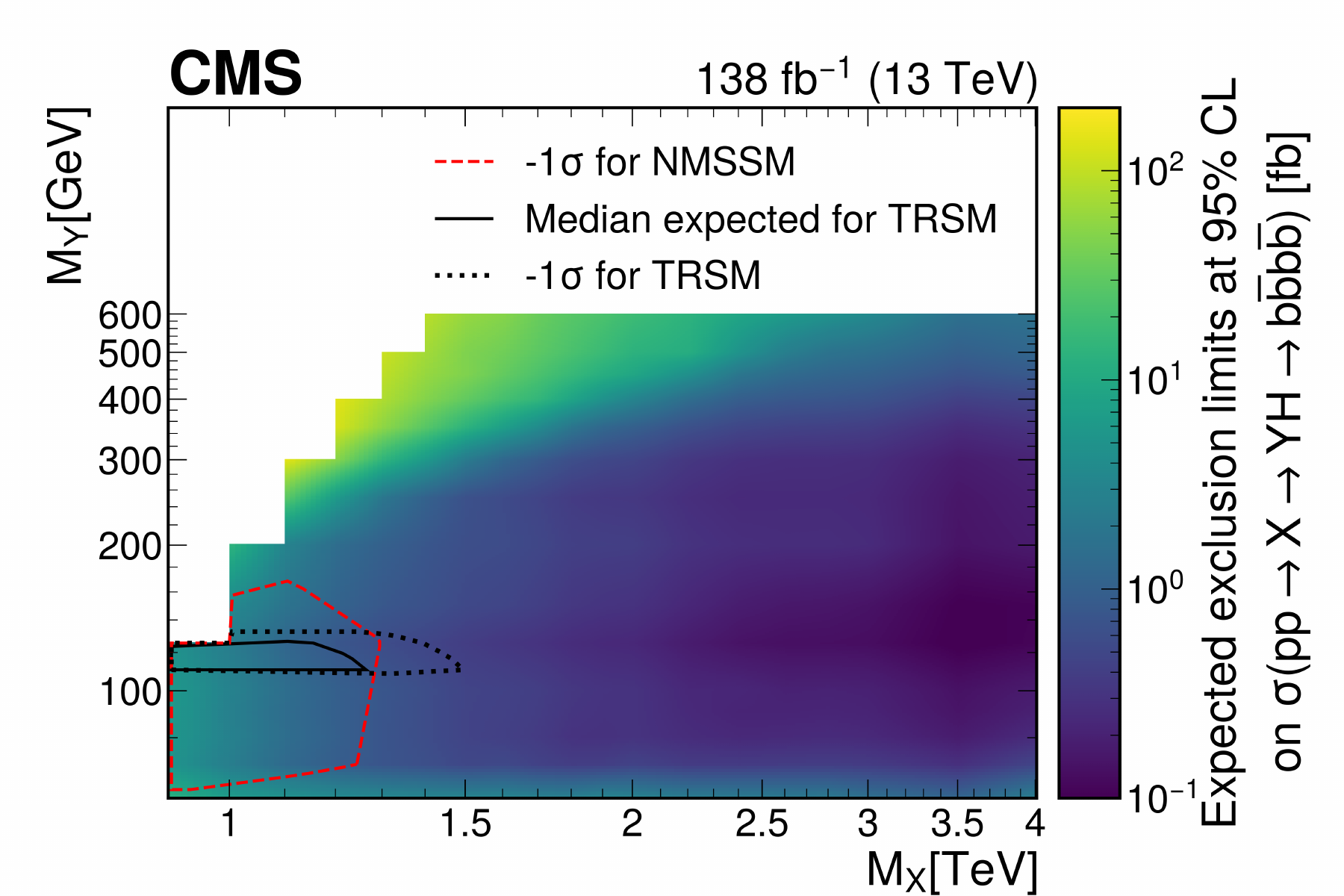}
  \includegraphics[width=0.49\textwidth]{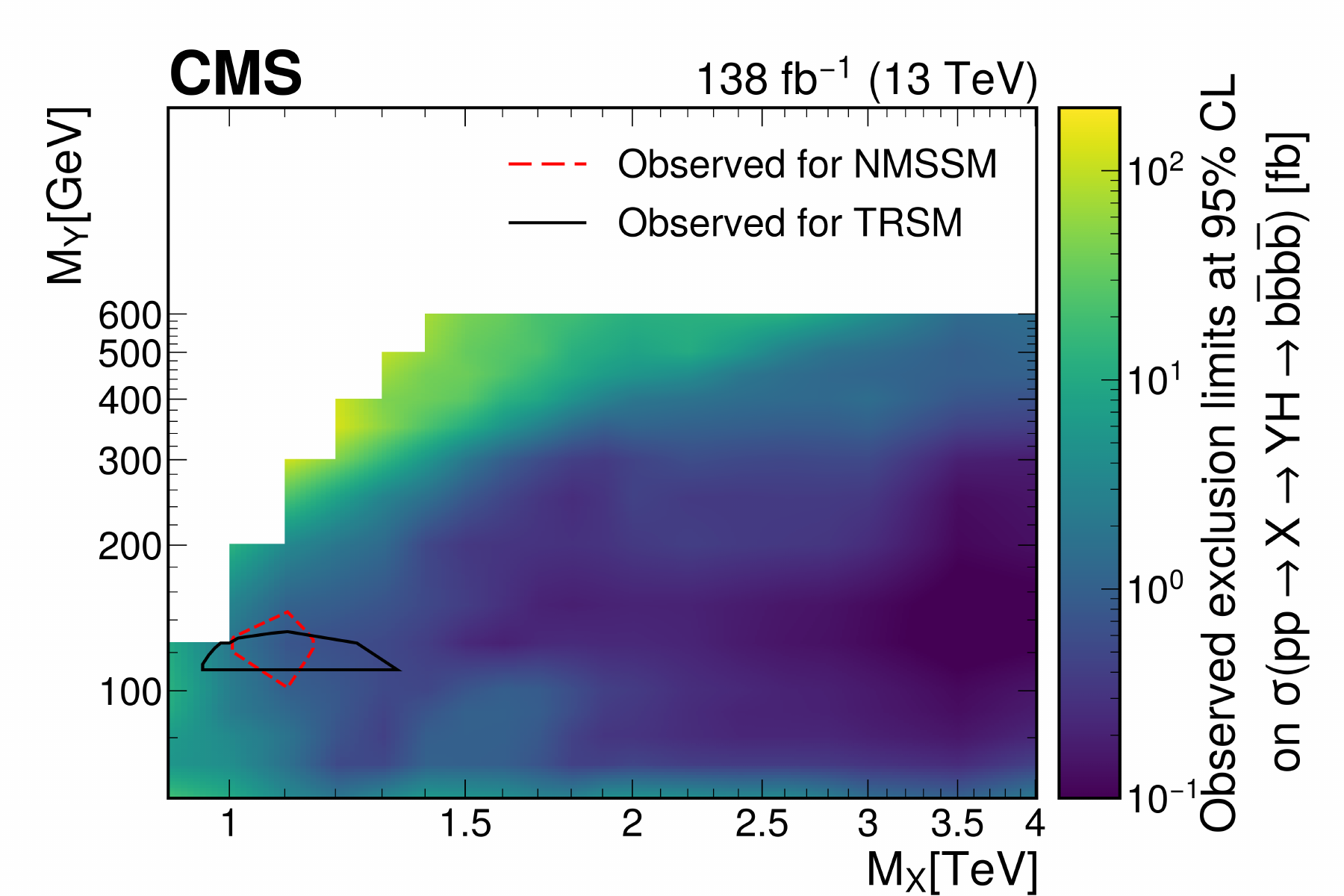}
  \caption{The 95\% confidence level expected (\cmsLeft) and observed (\cmsRight) upper limits on $\sigma(\Pp\Pp \to \XHbbYbb)$ for different values of \MX and \MY. The areas within the red and black contours represent the regions where the cross sections predicted by NMSSM and TRSM, respectively, are larger than the experimental limits. The areas within the dashed and dotted contours on the left show the excluded masses at $-1$ standard deviation from the expected limits.}
  \label{fig:lims2D}
\end{figure}

The  upper limits are computed with a modified frequentist approach, using the \CLs criterion~\cite{CLS1,CLS2} with the profile likelihood ratio used as the test-statistic and with the asymptotic approximation~\cite{Cowan2011}. As the signal distributions only assume that they originate from spin-0 particle decays, the limits are model-independent. The expected and observed limits at 95\% \CL as a function of \MX and \MY are shown in Fig.~\ref{fig:lims2D}, and range from 0.1\fb to 150\fb. 

The cross section limits are compared with the maximally allowed cross sections in the NMSSM and TRSM. In the NMSSM, no mass range is excluded by the median expected limits. However, the observed limits exclude an area with \MX range of 1.00--1.15\TeV and \MY range of 101--145\GeV.
For TRSM, an expected exclusion area with the bounds $0.90 < \MX < 1.26\TeV$ and $100 < \MY < 126\GeV$ is found while the observed exclusion range spans $0.95 < \MX < 1.33\TeV$ and $110 < \MY < 132\GeV$.

\section{Summary\label{sec:Summary}}

A search for massive scalar resonances \PX and \PY, where \PX decays to \PY and the standard model Higgs boson \PH, has been performed using proton-proton collision data collected at the LHC by the CMS detector between 2016 and 2018, and corresponding to an integrated luminosity of \intLumi. Events are selected using jet substructure and neural network based boosted $\PH/ \PY \to \bbbar$ identification algorithms. Upper limits at 95\% confidence level are set on the cross section of the process $\Pp\Pp \to \XHbbYbb$ for assumed masses of \PX in the range 0.9--4\TeV and \PY between 60--600\GeV. The expected and observed cross section limits for the considered process, set between 0.1 and 150\fb, are the  most stringent to date over much of the explored mass range.
These limits are interpreted as exclusion of possible \MX and \MY within the frameworks of the next-to-minimal supersymmetric model and the two-real-scalar-singlet extension of the standard model. 

\begin{acknowledgments}
We congratulate our colleagues in the CERN accelerator departments for the excellent performance of the LHC and thank the technical and administrative staffs at CERN and at other CMS institutes for their contributions to the success of the CMS effort. In addition, we gratefully acknowledge the computing centres and personnel of the Worldwide LHC Computing Grid and other centres for delivering so effectively the computing infrastructure essential to our analyses. Finally, we acknowledge the enduring support for the construction and operation of the LHC, the CMS detector, and the supporting computing infrastructure provided by the following funding agencies: BMBWF and FWF (Austria); FNRS and FWO (Belgium); CNPq, CAPES, FAPERJ, FAPERGS, and FAPESP (Brazil); MES and BNSF (Bulgaria); CERN; CAS, MoST, and NSFC (China); MINCIENCIAS (Colombia); MSES and CSF (Croatia); RIF (Cyprus); SENESCYT (Ecuador); MoER, ERC PUT and ERDF (Estonia); Academy of Finland, MEC, and HIP (Finland); CEA and CNRS/IN2P3 (France); BMBF, DFG, and HGF (Germany); GSRI (Greece); NKFIH (Hungary); DAE and DST (India); IPM (Iran); SFI (Ireland); INFN (Italy); MSIP and NRF (Republic of Korea); MES (Latvia); LAS (Lithuania); MOE and UM (Malaysia); BUAP, CINVESTAV, CONACYT, LNS, SEP, and UASLP-FAI (Mexico); MOS (Montenegro); MBIE (New Zealand); PAEC (Pakistan); MES and NSC (Poland); FCT (Portugal); MESTD (Serbia); MCIN/AEI and PCTI (Spain); MOSTR (Sri Lanka); Swiss Funding Agencies (Switzerland); MST (Taipei); MHESI and NSTDA (Thailand); TUBITAK and TENMAK (Turkey); NASU (Ukraine); STFC (United Kingdom); DOE and NSF (USA).
   
\hyphenation{Rachada-pisek} Individuals have received support from the Marie-Curie programme and the European Research Council and Horizon 2020 Grant, contract Nos.\ 675440, 724704, 752730, 758316, 765710, 824093, 884104, and COST Action CA16108 (European Union); the Leventis Foundation; the Alfred P.\ Sloan Foundation; the Alexander von Humboldt Foundation; the Belgian Federal Science Policy Office; the Fonds pour la Formation \`a la Recherche dans l'Industrie et dans l'Agriculture (FRIA-Belgium); the Agentschap voor Innovatie door Wetenschap en Technologie (IWT-Belgium); the F.R.S.-FNRS and FWO (Belgium) under the ``Excellence of Science -- EOS" -- be.h project n.\ 30820817; the Beijing Municipal Science \& Technology Commission, No. Z191100007219010; the Ministry of Education, Youth and Sports (MEYS) of the Czech Republic; the Hellenic Foundation for Research and Innovation (HFRI), Project Number 2288 (Greece); the Deutsche Forschungsgemeinschaft (DFG), under Germany's Excellence Strategy -- EXC 2121 ``Quantum Universe" -- 390833306, and under project number 400140256 - GRK2497; the Hungarian Academy of Sciences, the New National Excellence Program - \'UNKP, the NKFIH research grants K 124845, K 124850, K 128713, K 128786, K 129058, K 131991, K 133046, K 138136, K 143460, K 143477, 2020-2.2.1-ED-2021-00181, and TKP2021-NKTA-64 (Hungary); the Council of Science and Industrial Research, India; the Latvian Council of Science; the Ministry of Education and Science, project no. 2022/WK/14, and the National Science Center, contracts Opus 2021/41/B/ST2/01369 and 2021/43/B/ST2/01552 (Poland); the Funda\c{c}\~ao para a Ci\^encia e a Tecnologia, grant CEECIND/01334/2018 (Portugal); the National Priorities Research Program by Qatar National Research Fund; MCIN/AEI/10.13039/501100011033, ERDF ``a way of making Europe", and the Programa Estatal de Fomento de la Investigaci{\'o}n Cient{\'i}fica y T{\'e}cnica de Excelencia Mar\'{\i}a de Maeztu, grant MDM-2017-0765 and Programa Severo Ochoa del Principado de Asturias (Spain); the Chulalongkorn Academic into Its 2nd Century Project Advancement Project, and the National Science, Research and Innovation Fund via the Program Management Unit for Human Resources \& Institutional Development, Research and Innovation, grant B05F650021 (Thailand); the Kavli Foundation; the Nvidia Corporation; the SuperMicro Corporation; the Welch Foundation, contract C-1845; and the Weston Havens Foundation (USA).
\end{acknowledgments}

\bibliography{auto_generated} 
\cleardoublepage \appendix\section{The CMS Collaboration \label{app:collab}}\begin{sloppypar}\hyphenpenalty=5000\widowpenalty=500\clubpenalty=5000\input{B2G-21-003-public-authorlist.tex}\end{sloppypar}
\end{document}

%% file: B2G-21-003-public-authorlist.tex
\cmsinstitute{Yerevan Physics Institute, Yerevan, Armenia}
{\tolerance=6000
A.~Tumasyan\cmsAuthorMark{1}\cmsorcid{0009-0000-0684-6742}
\par}
\cmsinstitute{Institut f\"{u}r Hochenergiephysik, Vienna, Austria}
{\tolerance=6000
W.~Adam\cmsorcid{0000-0001-9099-4341}, J.W.~Andrejkovic, T.~Bergauer\cmsorcid{0000-0002-5786-0293}, S.~Chatterjee\cmsorcid{0000-0003-2660-0349}, K.~Damanakis\cmsorcid{0000-0001-5389-2872}, M.~Dragicevic\cmsorcid{0000-0003-1967-6783}, A.~Escalante~Del~Valle\cmsorcid{0000-0002-9702-6359}, R.~Fr\"{u}hwirth\cmsAuthorMark{2}\cmsorcid{0000-0002-0054-3369}, M.~Jeitler\cmsAuthorMark{2}\cmsorcid{0000-0002-5141-9560}, N.~Krammer\cmsorcid{0000-0002-0548-0985}, L.~Lechner\cmsorcid{0000-0002-3065-1141}, D.~Liko\cmsorcid{0000-0002-3380-473X}, I.~Mikulec\cmsorcid{0000-0003-0385-2746}, P.~Paulitsch, F.M.~Pitters, J.~Schieck\cmsAuthorMark{2}\cmsorcid{0000-0002-1058-8093}, R.~Sch\"{o}fbeck\cmsorcid{0000-0002-2332-8784}, D.~Schwarz\cmsorcid{0000-0002-3821-7331}, S.~Templ\cmsorcid{0000-0003-3137-5692}, W.~Waltenberger\cmsorcid{0000-0002-6215-7228}, C.-E.~Wulz\cmsAuthorMark{2}\cmsorcid{0000-0001-9226-5812}
% ...
\par}
\vskip\cmsinstskip
\dag:~Deceased\\
$^{1}$Also at Yerevan State University, Yerevan, Armenia\\
$^{2}$Also at TU Wien, Vienna, Austria\\
% ...


%% file: B2G-21-003_temp.bbl
\providecommand{\href}[2]{#2}\begingroup\raggedright\begin{thebibliography}{10}%
\makeatletter
\providecommand{\hrefCMSnoop }[0]{\@secondoftwo}%
\makeatother
\providecommand{\doi}{\texttt{doi:}\begingroup \urlstyle{tt}\Url}

\bibitem{HiggsDiscoveryAtlas}
\hrefCMSnoop {}{{ATLAS Collaboration}, ``{Observation of a new particle in the
  search for the standard model Higgs boson with the ATLAS detector at the
  LHC}'',} \textit{ Phys. Lett. B} \textbf{ 716} (2012) 01,
  \href{http://dx.doi.org/10.1016/j.physletb.2012.08.020}{\doi{10.1016/j.physletb.2012.08.020}},
\href{http://www.arXiv.org/abs/1207.7214}{\texttt{arXiv:1207.7214}}.

\bibitem{HiggsDiscoveryCMS}
\hrefCMSnoop {}{{CMS Collaboration}, ``{Observation of a new boson at a mass of
  125 GeV with the CMS experiment at the LHC}'',} \textit{ Phys. Lett. B}
  \textbf{ 716} (2012) 30,
  \href{http://dx.doi.org/10.1016/j.physletb.2012.08.021}{\doi{10.1016/j.physletb.2012.08.021}},
\href{http://www.arXiv.org/abs/1207.7235}{\texttt{arXiv:1207.7235}}.

\bibitem{CMSHiggsLongPaper}
\hrefCMSnoop {}{{CMS Collaboration}, ``{Observation of a new boson with mass
  near 125 GeV in pp collisions at $\sqrt{s}$ = 7 and 8 TeV}'',} \textit{ JHEP}
  \textbf{ 06} (2013) 081,
  \href{http://dx.doi.org/10.1007/JHEP06(2013)081}{\doi{10.1007/JHEP06(2013)081}},
\href{http://www.arXiv.org/abs/1303.4571}{\texttt{arXiv:1303.4571}}.

\bibitem{Englert:1964et}
\hrefCMSnoop {}{F.~Englert and R.~Brout, ``Broken symmetry and the mass of
  gauge vector mesons'',} \textit{ Phys. Rev. Lett.} \textbf{ 13} (1964) 321,
  \href{http://dx.doi.org/10.1103/PhysRevLett.13.321}{\doi{10.1103/PhysRevLett.13.321}}.

\bibitem{Higgs:1964ia}
\hrefCMSnoop {}{P.~W. Higgs, ``{Broken symmetries, massless particles and gauge
  fields}'',} \textit{ Phys. Lett.} \textbf{ 12} (1964) 132,
  \href{http://dx.doi.org/10.1016/0031-9163(64)91136-9}{\doi{10.1016/0031-9163(64)91136-9}}.

\bibitem{Higgs:1964pj}
\hrefCMSnoop {}{P.~W. Higgs, ``{Broken symmetries and the masses of gauge
  bosons}'',} \textit{ Phys. Rev. Lett.} \textbf{ 13} (1964) 508,
  \href{http://dx.doi.org/10.1103/PhysRevLett.13.508}{\doi{10.1103/PhysRevLett.13.508}}.

\bibitem{Guralnik:1964eu}
\hrefCMSnoop {}{G.~S. Guralnik, C.~R. Hagen, and T.~W.~B. Kibble, ``Global
  conservation laws and massless particles'',} \textit{ Phys. Rev. Lett.}
  \textbf{ 13} (1964) 585,
  \href{http://dx.doi.org/10.1103/PhysRevLett.13.585}{\doi{10.1103/PhysRevLett.13.585}}.

\bibitem{Higgs:1966ev}
\hrefCMSnoop {}{P.~W. Higgs, ``Spontaneous symmetry breakdown without massless
  bosons'',} \textit{ Phys. Rev.} \textbf{ 145} (1966) 1156,
  \href{http://dx.doi.org/10.1103/PhysRev.145.1156}{\doi{10.1103/PhysRev.145.1156}}.

\bibitem{Kibble:1967sv}
\hrefCMSnoop {}{T.~W.~B. Kibble, ``{Symmetry breaking in non-Abelian gauge
  theories}'',} \textit{ Phys. Rev.} \textbf{ 155} (1967) 1554,
  \href{http://dx.doi.org/10.1103/PhysRev.155.1554}{\doi{10.1103/PhysRev.155.1554}}.

\bibitem{Buttazzo:2013uya}
D.~Buttazzo\hrefCMSnoop {}{ {et~al.}, ``{Investigating the near-criticality of
  the Higgs boson}'',} \textit{ JHEP} \textbf{ 12} (2013) 089,
  \href{http://dx.doi.org/10.1007/JHEP12(2013)089}{\doi{10.1007/JHEP12(2013)089}},
  \href{http://www.arXiv.org/abs/1307.3536}{\texttt{arXiv:1307.3536}}.

\bibitem{Hamada:2012bp}
\hrefCMSnoop {}{Y.~Hamada, H.~Kawai, and K.-y. Oda, ``{Bare Higgs mass at
  Planck scale}'',} \textit{ Phys. Rev. D} \textbf{ 87} (2013) 053009,
  \href{http://dx.doi.org/10.1103/PhysRevD.87.053009}{\doi{10.1103/PhysRevD.87.053009}},
  \href{http://www.arXiv.org/abs/1210.2538}{\texttt{arXiv:1210.2538}}.
  [Erratum: \DOI{10.1103/PhysRevD.89.059901}].

\bibitem{Jones:2013aua}
\hrefCMSnoop {}{D.~R.~T. Jones, ``{Comment on \textquotedblleft{}Bare Higgs
  mass at Planck scale\textquotedblright{}}'',} \textit{ Phys. Rev. D} \textbf{
  88} (2013) 098301,
  \href{http://dx.doi.org/10.1103/PhysRevD.88.098301}{\doi{10.1103/PhysRevD.88.098301}},
  \href{http://www.arXiv.org/abs/1309.7335}{\texttt{arXiv:1309.7335}}.

\bibitem{Bednyakov:2015sca}
\hrefCMSnoop {}{A.~V. Bednyakov, B.~A. Kniehl, A.~F. Pikelner, and O.~L.
  Veretin, ``Stability of the electroweak vacuum: Gauge independence and
  advanced precision'',} \textit{ Phys. Rev. Lett.} \textbf{ 115} (2015)
  201802,
  \href{http://dx.doi.org/10.1103/PhysRevLett.115.201802}{\doi{10.1103/PhysRevLett.115.201802}},
  \href{http://www.arXiv.org/abs/1507.08833}{\texttt{arXiv:1507.08833}}.

\bibitem{WZSUSY}
\hrefCMSnoop {}{J.~Wess and B.~Zumino, ``Supergauge transformations in four
  dimensions'',} \textit{ Nucl. Phys. B} \textbf{ 70} (1974) 390,
  \href{http://dx.doi.org/10.1016/0550-3213(74)90355-1}{\doi{10.1016/0550-3213(74)90355-1}}.

\bibitem{RandallWED}
\hrefCMSnoop {}{L.~Randall and R.~Sundrum, ``{A large mass hierarchy from a
  small extra dimension}'',} \textit{ Phys. Rev. Lett.} \textbf{ 83} (1999)
  3370,
  \href{http://dx.doi.org/10.1103/PhysRevLett.83.3370}{\doi{10.1103/PhysRevLett.83.3370}},
\href{http://www.arXiv.org/abs/hep-ph/9905221}{\texttt{arXiv:hep-ph/9905221}}.

\bibitem{Fayet:1974pd}
\hrefCMSnoop {}{P.~Fayet, ``{Supergauge invariant extension of the Higgs
  mechanism and a model for the electron and its neutrino}'',} \textit{ Nucl.
  Phys. B} \textbf{ 90} (1975) 104,
  \href{http://dx.doi.org/10.1016/0550-3213(75)90636-7}{\doi{10.1016/0550-3213(75)90636-7}}.

\bibitem{Fayet:1977yc}
\hrefCMSnoop {}{P.~Fayet, ``{Spontaneously broken supersymmetric theories of
  weak, electromagnetic and strong interactions}'',} \textit{ Phys. Lett. B}
  \textbf{ 69} (1977) 489,
  \href{http://dx.doi.org/10.1016/0370-2693(77)90852-8}{\doi{10.1016/0370-2693(77)90852-8}}.

\bibitem{EllwangerNMSSM1}
\hrefCMSnoop {}{U.~Ellwanger, C.~Hugonie, and A.~M. Teixeira, ``{The
  next-to-minimal supersymmetric standard model}'',} \textit{ Phys. Rept.}
  \textbf{ 496} (2010) 1,
  \href{http://dx.doi.org/10.1016/j.physrep.2010.07.001}{\doi{10.1016/j.physrep.2010.07.001}},
  \href{http://www.arXiv.org/abs/0910.1785}{\texttt{arXiv:0910.1785}}.

\bibitem{ManiatisNMSSM2}
\hrefCMSnoop {}{M.~Maniatis, ``{The next-to-minimal supersymmetric extension of
  the standard model reviewed}'',} \textit{ Int. J. Mod. Phys. A} \textbf{ 25}
  (2010) 3505,
  \href{http://dx.doi.org/10.1142/S0217751X10049827}{\doi{10.1142/S0217751X10049827}},
  \href{http://www.arXiv.org/abs/0906.0777}{\texttt{arXiv:0906.0777}}.

\bibitem{KimMSSMmuProb}
\hrefCMSnoop {}{J.~E. Kim and H.~P. Nilles, ``{The $\mu$-problem and the strong
  CP problem}'',} \textit{ Phys. Lett. B} \textbf{ 138} (1984) 150,
  \href{http://dx.doi.org/10.1016/0370-2693(84)91890-2}{\doi{10.1016/0370-2693(84)91890-2}}.

\bibitem{ATLASHtotautau2020}
\hrefCMSnoop {}{{ATLAS Collaboration}, ``{Search for heavy Higgs bosons
  decaying into two tau leptons with the ATLAS detector using $pp$ collisions
  at $\sqrt{s}=13$ TeV}'',} \textit{ Phys. Rev. Lett.} \textbf{ 125} (2020)
  051801,
  \href{http://dx.doi.org/10.1103/PhysRevLett.125.051801}{\doi{10.1103/PhysRevLett.125.051801}},
  \href{http://www.arXiv.org/abs/2002.12223}{\texttt{arXiv:2002.12223}}.

\bibitem{CMSHtautau2018}
\hrefCMSnoop {}{{CMS Collaboration}, ``{Search for additional neutral MSSM
  Higgs bosons in the $\tau\tau$ final state in proton-proton collisions at
  $\sqrt{s}=$ 13 TeV}'',} \textit{ JHEP} \textbf{ 09} (2018) 007,
  \href{http://dx.doi.org/10.1007/JHEP09(2018)007}{\doi{10.1007/JHEP09(2018)007}},
  \href{http://www.arXiv.org/abs/1803.06553}{\texttt{arXiv:1803.06553}}.

\bibitem{Sirunyan:2684750}
\hrefCMSnoop {}{{CMS Collaboration}, ``{Search for heavy Higgs bosons decaying
  to a top quark pair in proton-proton collisions at $\sqrt{s} = $ 13 TeV}'',}
  \textit{ JHEP} \textbf{ 04} (2020) 171,
  \href{http://dx.doi.org/10.1007/JHEP04(2020)171}{\doi{10.1007/JHEP04(2020)171}},
  \href{http://www.arXiv.org/abs/1908.01115}{\texttt{arXiv:1908.01115}}.

\bibitem{Djouadi:2008uw}
A.~Djouadi\hrefCMSnoop {}{ {et~al.}, ``{Benchmark scenarios for the NMSSM}'',}
  \textit{ JHEP} \textbf{ 07} (2008) 002,
  \href{http://dx.doi.org/10.1088/1126-6708/2008/07/002}{\doi{10.1088/1126-6708/2008/07/002}},
  \href{http://www.arXiv.org/abs/0801.4321}{\texttt{arXiv:0801.4321}}.

\bibitem{Ellwanger:2017skc}
\hrefCMSnoop {}{U.~Ellwanger and M.~Rodriguez-Vazquez, ``{Simultaneous search
  for extra light and heavy Higgs bosons via cascade decays}'',} \textit{ JHEP}
  \textbf{ 11} (2017) 008,
  \href{http://dx.doi.org/10.1007/JHEP11(2017)008}{\doi{10.1007/JHEP11(2017)008}},
  \href{http://www.arXiv.org/abs/1707.08522}{\texttt{arXiv:1707.08522}}.

\bibitem{Ellwanger2022}
\hrefCMSnoop {}{U.~Ellwanger and C.~Hugonie, ``Benchmark planes for
  higgs-to-higgs decays in the {NMSSM}'',} \textit{ Eur. Phys. J. C} \textbf{
  82} (2022), no.~5,
  \href{http://dx.doi.org/10.1140/epjc/s10052-022-10364-3}{\doi{10.1140/epjc/s10052-022-10364-3}},
  \href{http://www.arXiv.org/abs/2203.05049}{\texttt{arXiv:2203.05049}}.

\bibitem{CMSXtoHYtotautaubb2021}
\hrefCMSnoop {}{{CMS Collaboration}, ``Search for a heavy {Higgs} boson
  decaying into two lighter {Higgs} bosons in the $\uptau$$\uptau$bb final
  state at 13 {TeV}'',} \textit{ JHEP} \textbf{ 11} (2021) 057,
  \href{http://dx.doi.org/10.1007/jhep11(2021)057}{\doi{10.1007/jhep11(2021)057}},
  \href{http://www.arXiv.org/abs/2106.10361}{\texttt{arXiv:2106.10361}}.

\bibitem{Robens:2019kga}
\hrefCMSnoop {}{T.~Robens, T.~Stefaniak, and J.~Wittbrodt,
  ``{Two-real-scalar-singlet extension of the SM: LHC phenomenology and
  benchmark scenarios}'',} \textit{ Eur. Phys. J. C} \textbf{ 80} (2020) 151,
  \href{http://dx.doi.org/10.1140/epjc/s10052-020-7655-x}{\doi{10.1140/epjc/s10052-020-7655-x}},
  \href{http://www.arXiv.org/abs/1908.08554}{\texttt{arXiv:1908.08554}}.

\bibitem{Sirunyan:2759951}
\hrefCMSnoop {}{{CMS Collaboration}, ``{Precision luminosity measurement in
  proton-proton collisions at $\sqrt{s} =$ 13 TeV in 2015 and 2016 at CMS}'',}
  \textit{ Eur. Phys. J. C} \textbf{ 81} (2021) 800,
  \href{http://dx.doi.org/10.1140/epjc/s10052-021-09538-2}{\doi{10.1140/epjc/s10052-021-09538-2}},
  \href{http://www.arXiv.org/abs/2104.01927}{\texttt{arXiv:2104.01927}}.

\bibitem{CMS-PAS-LUM-17-004}
\href {https://cds.cern.ch/record/2621960/}{{CMS Collaboration}, ``{CMS}
  luminosity measurement for the 2017 data-taking period at $\sqrt{s}$ = 13
  {TeV}'',} CMS Physics Analysis Summary CMS-PAS-LUM-17-004, 2018.

\bibitem{CMS-PAS-LUM-18-002}
\href {https://cds.cern.ch/record/2676164/}{{CMS Collaboration}, ``{CMS}
  luminosity measurement for the 2018 data-taking period at $\sqrt{s}$ = 13
  {TeV}'',} CMS Physics Analysis Summary CMS-PAS-LUM-18-002, 2019.

\bibitem{hepdata}
\hrefCMSnoop {}{}{HEPD}ata record for this analysis, 2022.
\newblock
  \href{http://dx.doi.org/10.17182/hepdata.115995}{\doi{10.17182/hepdata.115995}}.

\bibitem{CMSJINST}
\hrefCMSnoop {}{{CMS Collaboration}, ``The {CMS} experiment at the {CERN}
  {LHC}'',} \textit{ JINST} \textbf{ 3} (2008) S08004,
  \href{http://dx.doi.org/10.1088/1748-0221/3/08/S08004}{\doi{10.1088/1748-0221/3/08/S08004}}.

\bibitem{CMSL1}
\hrefCMSnoop {}{{CMS Collaboration}, ``{Performance of the CMS Level-1 trigger
  in proton-proton collisions at $\sqrt{s} = 13$\,TeV}'',} \textit{ JINST}
  \textbf{ 15} (2020) P10017,
  \href{http://dx.doi.org/10.1088/1748-0221/15/10/P10017}{\doi{10.1088/1748-0221/15/10/P10017}},
  \href{http://www.arXiv.org/abs/2006.10165}{\texttt{arXiv:2006.10165}}.

\bibitem{CMSHLT}
\hrefCMSnoop {}{{CMS Collaboration}, ``{The CMS trigger system}'',} \textit{
  JINST} \textbf{ 12} (2017) P01020,
  \href{http://dx.doi.org/10.1088/1748-0221/12/01/P01020}{\doi{10.1088/1748-0221/12/01/P01020}},
\href{http://www.arXiv.org/abs/1609.02366}{\texttt{arXiv:1609.02366}}.

\bibitem{CMS-TDR-15-02}
\href {http://cds.cern.ch/record/2020886}{{CMS Collaboration}, ``Technical
  proposal for the {Phase-II} upgrade of the {Compact Muon Solenoid}'',} CMS
  Technical Proposal CERN-LHCC-2015-010, CMS-TDR-15-02, 2015.

\bibitem{CMSPF}
\hrefCMSnoop {}{{CMS Collaboration}, ``{Particle-flow reconstruction and global
  event description with the CMS detector}'',} \textit{ JINST} \textbf{ 12}
  (2017) P10003,
  \href{http://dx.doi.org/10.1088/1748-0221/12/10/P10003}{\doi{10.1088/1748-0221/12/10/P10003}},
  \href{http://www.arXiv.org/abs/1706.04965}{\texttt{arXiv:1706.04965}}.

\bibitem{Cacciari:2008gp}
\hrefCMSnoop {}{M.~Cacciari, G.~P. Salam, and G.~Soyez, ``{The anti-\kt jet
  clustering algorithm}'',} \textit{ JHEP} \textbf{ 04} (2008) 063,
  \href{http://dx.doi.org/10.1088/1126-6708/2008/04/063}{\doi{10.1088/1126-6708/2008/04/063}},
  \href{http://www.arXiv.org/abs/0802.1189}{\texttt{arXiv:0802.1189}}.

\bibitem{Cacciari:2011ma}
\hrefCMSnoop {}{M.~Cacciari, G.~P. Salam, and G.~Soyez, ``{FastJet user
  manual}'',} \textit{ Eur. Phys. J. C} \textbf{ 72} (2012) 1896,
  \href{http://dx.doi.org/10.1140/epjc/s10052-012-1896-2}{\doi{10.1140/epjc/s10052-012-1896-2}},
\href{http://www.arXiv.org/abs/1111.6097}{\texttt{arXiv:1111.6097}}.

\bibitem{CMS:2016lmd}
\hrefCMSnoop {}{{CMS Collaboration}, ``{Jet energy scale and resolution in the
  CMS experiment in pp collisions at 8 TeV}'',} \textit{ JINST} \textbf{ 12}
  (2017) P02014,
  \href{http://dx.doi.org/10.1088/1748-0221/12/02/P02014}{\doi{10.1088/1748-0221/12/02/P02014}},
\href{http://www.arXiv.org/abs/1607.03663}{\texttt{arXiv:1607.03663}}.

\bibitem{Sirunyan:2020foa}
\hrefCMSnoop {}{{CMS Collaboration}, ``{Pileup mitigation at CMS in 13 TeV
  data}'',} \textit{ JINST} \textbf{ 15} (2020) P09018,
  \href{http://dx.doi.org/10.1088/1748-0221/15/09/p09018}{\doi{10.1088/1748-0221/15/09/p09018}},
  \href{http://www.arXiv.org/abs/2003.00503}{\texttt{arXiv:2003.00503}}.

\bibitem{SalamJetography}
\hrefCMSnoop {}{G.~P. Salam, ``Towards jetography'',} \textit{ Eur. Phys. J. C}
  \textbf{ 67} (2010) 637,
  \href{http://dx.doi.org/10.1140/epjc/s10052-010-1314-6}{\doi{10.1140/epjc/s10052-010-1314-6}},
\href{http://www.arXiv.org/abs/0906.1833}{\texttt{arXiv:0906.1833}}.

\bibitem{Krohn:2009th}
\hrefCMSnoop {}{D.~Krohn, J.~Thaler, and L.-T. Wang, ``Jet trimming'',}
  \textit{ JHEP} \textbf{ 02} (2010) 084,
  \href{http://dx.doi.org/10.1007/JHEP02(2010)084}{\doi{10.1007/JHEP02(2010)084}},
\href{http://www.arXiv.org/abs/0912.1342}{\texttt{arXiv:0912.1342}}.

\bibitem{Dasgupta:2013ihk}
\hrefCMSnoop {}{M.~Dasgupta, A.~Fregoso, S.~Marzani, and G.~P. Salam,
  ``{Towards an understanding of jet substructure}'',} \textit{ JHEP} \textbf{
  09} (2013) 029,
  \href{http://dx.doi.org/10.1007/JHEP09(2013)029}{\doi{10.1007/JHEP09(2013)029}},
\href{http://www.arXiv.org/abs/1307.0007}{\texttt{arXiv:1307.0007}}.

\bibitem{Larkoski:2014wba}
\hrefCMSnoop {}{A.~J. Larkoski, S.~Marzani, G.~Soyez, and J.~Thaler, ``Soft
  drop'',} \textit{ JHEP} \textbf{ 05} (2014) 146,
  \href{http://dx.doi.org/10.1007/JHEP05(2014)146}{\doi{10.1007/JHEP05(2014)146}},
\href{http://www.arXiv.org/abs/1402.2657}{\texttt{arXiv:1402.2657}}.

\bibitem{Bertolini:2014bba}
\hrefCMSnoop {}{D.~Bertolini, P.~Harris, M.~Low, and N.~Tran, ``Pileup per
  particle identification'',} \textit{ JHEP} \textbf{ 10} (2014) 059,
  \href{http://dx.doi.org/10.1007/JHEP10(2014)059}{\doi{10.1007/JHEP10(2014)059}},
\href{http://www.arXiv.org/abs/1407.6013}{\texttt{arXiv:1407.6013}}.

\bibitem{CMSMET}
\hrefCMSnoop {}{{CMS Collaboration}, ``Performance of missing transverse
  momentum reconstruction in proton-proton collisions at $\sqrt{s} = 13$\,{TeV}
  using the {CMS} detector'',} \textit{ JINST} \textbf{ 14} (2019) P07004,
  \href{http://dx.doi.org/10.1088/1748-0221/14/07/P07004}{\doi{10.1088/1748-0221/14/07/P07004}},
\href{http://www.arXiv.org/abs/1903.06078}{\texttt{arXiv:1903.06078}}.

\bibitem{Madgraph}
J.~Alwall\hrefCMSnoop {}{ {et~al.}, ``The automated computation of tree-level
  and next-to-leading order differential cross sections, and their matching to
  parton shower simulations'',} \textit{ JHEP} \textbf{ 07} (2014) 079,
  \href{http://dx.doi.org/10.1007/JHEP07(2014)079}{\doi{10.1007/JHEP07(2014)079}},
  \href{http://www.arXiv.org/abs/1405.0301}{\texttt{arXiv:1405.0301}}.

\bibitem{NMSSMufo}
D.~Curtin\hrefCMSnoop {}{ {et~al.}, ``Exotic decays of the 125 {GeV} {Higgs}
  boson'',} \textit{ Phys. Rev. D} \textbf{ 90} (2014) 075004,
  \href{http://dx.doi.org/10.1103/physrevd.90.075004}{\doi{10.1103/physrevd.90.075004}},
  \href{http://www.arXiv.org/abs/1312.4992}{\texttt{arXiv:1312.4992}}.

\bibitem{UFO}
A.~Alloul\hrefCMSnoop {}{ {et~al.}, ``{FeynRules 2.0 --- a complete toolbox for
  tree-level phenomenology}'',} \textit{ Comput. Phys. Commun.} \textbf{ 185}
  (2014) 2250,
  \href{http://dx.doi.org/10.1016/j.cpc.2014.04.012}{\doi{10.1016/j.cpc.2014.04.012}},
  \href{http://www.arXiv.org/abs/1310.1921}{\texttt{arXiv:1310.1921}}.

\bibitem{POWHEG}
\hrefCMSnoop {}{S.~Frixione, G.~Ridolfi, and P.~Nason, ``A positive-weight
  next-to-leading-order {Monte Carlo} for heavy flavour hadroproduction'',}
  \textit{ JHEP} \textbf{ 09} (2007) 126,
  \href{http://dx.doi.org/10.1088/1126-6708/2007/09/126}{\doi{10.1088/1126-6708/2007/09/126}},
  \href{http://www.arXiv.org/abs/0707.3088}{\texttt{arXiv:0707.3088}}.

\bibitem{POWHEG_Frixione}
\hrefCMSnoop {}{S.~Frixione, P.~Nason, and C.~Oleari, ``Matching {NLO} {QCD}
  computations with parton shower simulations: the {POWHEG} method'',} \textit{
  JHEP} \textbf{ 11} (2007) 070,
  \href{http://dx.doi.org/10.1088/1126-6708/2007/11/070}{\doi{10.1088/1126-6708/2007/11/070}},
\href{http://www.arXiv.org/abs/0709.2092}{\texttt{arXiv:0709.2092}}.

\bibitem{POWHEGBOX}
\hrefCMSnoop {}{S.~Alioli, P.~Nason, C.~Oleari, and E.~Re, ``{A general
  framework for implementing NLO calculations in shower Monte Carlo programs:
  the POWHEG BOX}'',} \textit{ JHEP} \textbf{ 06} (2010) 043,
  \href{http://dx.doi.org/10.1007/JHEP06(2010)043}{\doi{10.1007/JHEP06(2010)043}},
\href{http://www.arXiv.org/abs/1002.2581}{\texttt{arXiv:1002.2581}}.

\bibitem{POWHEG_Nason}
\hrefCMSnoop {}{P.~Nason, ``A new method for combining {NLO} {QCD} with shower
  monte carlo algorithms'',} \textit{ JHEP} \textbf{ 11} (2004) 040,
  \href{http://dx.doi.org/10.1088/1126-6708/2004/11/040}{\doi{10.1088/1126-6708/2004/11/040}},
  \href{http://www.arXiv.org/abs/hep-ph/0409146}{\texttt{arXiv:hep-ph/0409146}}.

\bibitem{Czakon:2011xx}
\hrefCMSnoop {}{M.~Czakon and A.~Mitov, ``Top++: {A} program for the
  calculation of the top-pair cross-section at hadron colliders'',} \textit{
  Comput. Phys. Commun.} \textbf{ 185} (2014) 2930,
  \href{http://dx.doi.org/10.1016/j.cpc.2014.06.021}{\doi{10.1016/j.cpc.2014.06.021}},
\href{http://www.arXiv.org/abs/1112.5675}{\texttt{arXiv:1112.5675}}.

\bibitem{NNPDF31}
\hrefCMSnoop {}{{NNPDF} Collaboration, ``{Parton distributions from
  high-precision collider data}'',} \textit{ Eur. Phys. J. C} \textbf{ 77}
  (2017) 663,
  \href{http://dx.doi.org/10.1140/epjc/s10052-017-5199-5}{\doi{10.1140/epjc/s10052-017-5199-5}},
  \href{http://www.arXiv.org/abs/1706.00428}{\texttt{arXiv:1706.00428}}.

\bibitem{Buckley:2014ana}
A.~Buckley\hrefCMSnoop {}{ {et~al.}, ``{LHAPDF6: parton density access in the
  LHC precision era}'',} \textit{ Eur. Phys. J. C} \textbf{ 75} (2015) 132,
  \href{http://dx.doi.org/10.1140/epjc/s10052-015-3318-8}{\doi{10.1140/epjc/s10052-015-3318-8}},
\href{http://www.arXiv.org/abs/1412.7420}{\texttt{arXiv:1412.7420}}.

\bibitem{NNPDF30}
\hrefCMSnoop {}{{NNPDF} Collaboration, ``Parton distributions for the {LHC Run
  II}'',} \textit{ JHEP} \textbf{ 04} (2015) 040,
  \href{http://dx.doi.org/10.1007/JHEP04(2015)040}{\doi{10.1007/JHEP04(2015)040}},
\href{http://www.arXiv.org/abs/1410.8849}{\texttt{arXiv:1410.8849}}.

\bibitem{Pythia8p2}
T.~Sj{\"{o}}strand\hrefCMSnoop {}{ {et~al.}, ``{An introduction to PYTHIA
  8.2}'',} \textit{ Comput. Phys. Commun.} \textbf{ 191} (2015) 159,
  \href{http://dx.doi.org/10.1016/j.cpc.2015.01.024}{\doi{10.1016/j.cpc.2015.01.024}},
\href{http://www.arXiv.org/abs/1410.3012}{\texttt{arXiv:1410.3012}}.

\bibitem{MLM}
J.~Alwall\hrefCMSnoop {}{ {et~al.}, ``Comparative study of various algorithms
  for the merging of parton showers and matrix elements in hadronic
  collisions'',} \textit{ Eur. Phys. J. C} \textbf{ 53} (2007) 473,
  \href{http://dx.doi.org/10.1140/epjc/s10052-007-0490-5}{\doi{10.1140/epjc/s10052-007-0490-5}},
  \href{http://www.arXiv.org/abs/0706.2569}{\texttt{arXiv:0706.2569}}.

\bibitem{CP5}
\hrefCMSnoop {}{{CMS Collaboration}, ``{Extraction and validation of a new set
  of CMS PYTHIA8 tunes from underlying-event measurements}'',} \textit{ Eur.
  Phys. J. C} \textbf{ 80} (2020) 4,
  \href{http://dx.doi.org/10.1140/epjc/s10052-019-7499-4}{\doi{10.1140/epjc/s10052-019-7499-4}},
  \href{http://www.arXiv.org/abs/1903.12179}{\texttt{arXiv:1903.12179}}.

\bibitem{CUETP8M2T4}
\href {https://cds.cern.ch/record/2235192}{{CMS Collaboration},
  ``{Investigations of the impact of the parton shower tuning in Pythia 8 in
  the modelling of $\mathrm{t\overline{t}}$ at $\sqrt{s}=8$ and 13 TeV}'',}
  {CMS Physics Analysis Summary} CMS-PAS-TOP-16-021, 2016.

\bibitem{CUETP8M1}
\hrefCMSnoop {}{{CMS Collaboration}, ``{Event generator tunes obtained from
  underlying event and multiparton scattering measurements}'',} \textit{ Eur.
  Phys. J. C} \textbf{ 76} (2016) 155,
  \href{http://dx.doi.org/10.1140/epjc/s10052-016-3988-x}{\doi{10.1140/epjc/s10052-016-3988-x}},
  \href{http://www.arXiv.org/abs/1512.00815}{\texttt{arXiv:1512.00815}}.

\bibitem{GEANT}
\hrefCMSnoop {}{J.~Allison {et~al.}, ``{\GEANTfour developments and
  applications}'',} \textit{ IEEE Trans. Nucl. Sci.} \textbf{ 53} (2006) 270,
\href{http://dx.doi.org/10.1109/TNS.2006.869826}{\doi{10.1109/TNS.2006.869826}}.

\bibitem{CMSLHCppXsec}
\hrefCMSnoop {}{{CMS Collaboration}, ``{Measurement of the inelastic
  proton-proton cross section at $\sqrt{s} = 13\TeV$}'',} \textit{ JHEP}
  \textbf{ 07} (2018) 161,
  \href{http://dx.doi.org/10.1007/JHEP07(2018)161}{\doi{10.1007/JHEP07(2018)161}},
\href{http://www.arXiv.org/abs/1802.02613}{\texttt{arXiv:1802.02613}}.

\bibitem{CMSBTagging2018}
\hrefCMSnoop {}{{CMS Collaboration}, ``{Identification of heavy-flavour jets
  with the CMS detector in pp collisions at 13 TeV}'',} \textit{ JINST}
  \textbf{ 13} (2018) P05011,
  \href{http://dx.doi.org/10.1088/1748-0221/13/05/P05011}{\doi{10.1088/1748-0221/13/05/P05011}},
  \href{http://www.arXiv.org/abs/1712.07158}{\texttt{arXiv:1712.07158}}.

\bibitem{PNet_HQu_LGouskos}
\hrefCMSnoop {}{H.~Qu and L.~Gouskos, ``{Jet tagging via particle clouds}'',}
  \textit{ Phys. Rev. D} \textbf{ 101} (2020) 056019,
  \href{http://dx.doi.org/10.1103/PhysRevD.101.056019}{\doi{10.1103/PhysRevD.101.056019}},
  \href{http://www.arXiv.org/abs/1902.08570}{\texttt{arXiv:1902.08570}}.

\bibitem{CMS-DP-2020-002}
\href {https://cds.cern.ch/record/2707946}{{CMS Collaboration},
  ``{Identification of highly Lorentz-boosted heavy particles using graph
  neural networks and new mass decorrelation techniques}'',} {CMS Detector
  Performance Report} CMS-DP-2020-002, 2020.

\bibitem{CMS:2015xaf}
\hrefCMSnoop {}{{CMS Collaboration}, ``{Performance of electron reconstruction
  and selection with the CMS detector in proton-proton collisions at $\sqrt{s}
  = 8$\TeV}'',} \textit{ JINST} \textbf{ 10} (2015) P06005,
  \href{http://dx.doi.org/10.1088/1748-0221/10/06/P06005}{\doi{10.1088/1748-0221/10/06/P06005}},
\href{http://www.arXiv.org/abs/1502.02701}{\texttt{arXiv:1502.02701}}.

\bibitem{CMS:2018rym}
\hrefCMSnoop {}{{CMS Collaboration}, ``{Performance of the CMS muon detector
  and muon reconstruction with proton-proton collisions at $\sqrt{s}=$ 13
  TeV}'',} \textit{ JINST} \textbf{ 13} (2018) P06015,
  \href{http://dx.doi.org/10.1088/1748-0221/13/06/P06015}{\doi{10.1088/1748-0221/13/06/P06015}},
  \href{http://www.arXiv.org/abs/1804.04528}{\texttt{arXiv:1804.04528}}.

\bibitem{DeepJet}
E.~Bols\hrefCMSnoop {}{ {et~al.}, ``{Jet flavour classification using
  DeepJet}'',} \textit{ JINST} \textbf{ 15} (2020) P12012,
  \href{http://dx.doi.org/10.1088/1748-0221/15/12/P12012}{\doi{10.1088/1748-0221/15/12/P12012}},
  \href{http://www.arXiv.org/abs/2008.10519}{\texttt{arXiv:2008.10519}}.

\bibitem{Fisher}
R.~G. Lomax and D.~L. Hahs-Vaughn, ``{Statistical concepts: a second course}''.
\newblock Taylor and Francis, 2012.

\bibitem{CMS-PAS-TOP-17-007}
\hrefCMSnoop {}{{CMS Collaboration}, ``{Measurement of the top quark mass with
  lepton+jets final states using $\Pp\Pp$ collisions at $\sqrt{s} =
  13\TeV$}'',} \textit{ Eur. Phys. J. C} \textbf{ 78} (2018) 891,
  \href{http://dx.doi.org/10.1140/epjc/s10052-018-6332-9}{\doi{10.1140/epjc/s10052-018-6332-9}},
\href{http://www.arXiv.org/abs/1805.01428}{\texttt{arXiv:1805.01428}}.

\bibitem{HHbbbb2016Paper}
\hrefCMSnoop {}{{CMS Collaboration}, ``Search for a massive resonance decaying
  to a pair of {Higgs} bosons in the four b quark final state in
  proton{\textendash}proton collisions at $\sqrt{s}$ = 13 {TeV}'',} \textit{
  Physics Letters B} \textbf{ 781} (2018) 244,
  \href{http://dx.doi.org/10.1016/j.physletb.2018.03.084}{\doi{10.1016/j.physletb.2018.03.084}},
  \href{http://www.arXiv.org/abs/1710.04960}{\texttt{arXiv:1710.04960}}.

\bibitem{Butterworth:2015oua}
\hrefCMSnoop {}{J.~Butterworth {et~al.}, ``{PDF4LHC} recommendations for {LHC
  Run II}'',} \textit{ J. Phys. G} \textbf{ 43} (2016) 023001,
  \href{http://dx.doi.org/10.1088/0954-3899/43/2/023001}{\doi{10.1088/0954-3899/43/2/023001}},
\href{http://www.arXiv.org/abs/1510.03865}{\texttt{arXiv:1510.03865}}.

\bibitem{BarlowBeeston}
\hrefCMSnoop {}{R.~Barlow and C.~Beeston, ``Fitting using finite {Monte Carlo}
  samples'',} \textit{ Comput. Phys. Commun.} \textbf{ 77} (1993) 219,
  \href{http://dx.doi.org/10.1016/0010-4655(93)90005-W}{\doi{10.1016/0010-4655(93)90005-W}}.

\bibitem{BBLite}
\hrefCMSnoop {}{J.~S. Conway, ``Incorporating nuisance parameters in
  likelihoods for multisource spectra'',} in \textit{ {P}roceedings, workshop
  on statistical issues related to discovery claims in search experiments and
  unfolding ({PHYSTAT} 2011)}.
\newblock 2011.
\newblock \href{http://www.arXiv.org/abs/1103.0354}{\texttt{arXiv:1103.0354}}.
\newblock
  \href{http://dx.doi.org/10.5170/CERN-2011-006.115}{\doi{10.5170/CERN-2011-006.115}}.

\bibitem{Vitells2011}
\hrefCMSnoop {}{O.~Vitells and E.~Gross, ``{Estimating the significance of a
  signal in a multi-dimensional search}'',} \textit{ Astropart. Phys.} \textbf{
  35} (2011) 230,
  \href{http://dx.doi.org/10.1016/j.astropartphys.2011.08.005}{\doi{10.1016/j.astropartphys.2011.08.005}},
  \href{http://www.arXiv.org/abs/1105.4355}{\texttt{arXiv:1105.4355}}.

\bibitem{CLS1}
\hrefCMSnoop {}{A.~L. Read, ``{Presentation of search results: the CL$_s$
  technique}'',} \textit{ J. Phys. G} \textbf{ 28} (2002) 2693,
  \href{http://dx.doi.org/10.1088/0954-3899/28/10/313}{\doi{10.1088/0954-3899/28/10/313}}.

\bibitem{CLS2}
\hrefCMSnoop {}{T.~Junk, ``{Confidence level computation for combining searches
  with small statistics}'',} \textit{ Nucl. Instrum. Meth. A} \textbf{ 434}
  (1999) 435,
  \href{http://dx.doi.org/10.1016/S0168-9002(99)00498-2}{\doi{10.1016/S0168-9002(99)00498-2}},
  \href{http://www.arXiv.org/abs/hep-ex/9902006}{\texttt{arXiv:hep-ex/9902006}}.

\bibitem{Cowan2011}
\hrefCMSnoop {}{G.~Cowan, K.~Cranmer, E.~Gross, and O.~Vitells, ``Asymptotic
  formulae for likelihood-based tests of new physics'',} \textit{ Eur. Phys. J.
  C} \textbf{ 71} (2011) 1554,
  \href{http://dx.doi.org/10.1140/epjc/s10052-011-1554-0}{\doi{10.1140/epjc/s10052-011-1554-0}},
  \href{http://www.arXiv.org/abs/1007.1727}{\texttt{arXiv:1007.1727}}.
[Erratum: \DOI{10.1140/epjc/s10052-013-2501-z}].

\end{thebibliography}\endgroup
